\documentstyle[12pt]{article}

\def\hybrid{\topmargin -20pt  \oddsidemargin 0pt
      \headheight 0pt   \headsep 0pt
      \textwidth 6.25in 
      \textheight 9.5in 
      \marginparwidth .875in
      \parskip 5pt plus 1pt   \jot = 1.5ex}

\hybrid

\def\x{\times}
\def\ox{\otimes}
\def\o+{\oplus}
\def\ra{\rightarrow}
\def\lra{\longrightarrow}
\def\Lra{\Longrightarrow}

\def\Llra{\Longleftrightarrow}

\def\beqa{\begin{eqnarray}}
\def\eeqa{\end{eqnarray}}
                       
\newcommand{\ov}{\overline}
\newcommand{\un}{\underline}
\newcommand{\p}{\partial}

\newcommand{\al}{\alpha}

\newcommand{\ga}{\gamma}

\newcommand{\la}{\lambda}
\newcommand{\si}{\sigma}

\newcommand{\A}{{\cal A}}
\newcommand{\C}{{\cal C}}
\newcommand{\E}{{\cal E}}
\newcommand{\cL}{{\cal L}}
\newcommand{\M}{{\cal M}}
\newcommand{\N}{{\cal N}}
\newcommand{\cO}{{\cal O}}
\newcommand{\cP}{{\cal P}}
\newcommand{\R}{{\cal R}}
\newcommand{\cS}{{\cal S}}
\def\Si{\Sigma}

\newcommand{\resetcounter}{\setcounter{equation}{0}}

\begin{document}
\thispagestyle{empty}
\rightline{LMU-ASC  34/10}
\vspace{2truecm}
\centerline{\bf \LARGE On the Heterotic World-sheet Instanton Superpotential}
\vspace{.5truecm}
\centerline{\bf \LARGE and its individual Contributions}

\vspace{1.5truecm}
\centerline{Gottfried 
Curio\footnote{gottfried.curio@physik.uni-muenchen.de}$^{,}$\footnote{supported
by DFG grant CU 191/1-1}} 

\vspace{.6truecm}

\centerline{{\em Arnold-Sommerfeld-Center for Theoretical Physics}}
\centerline{{\em Department f\"ur Physik, 
Ludwig-Maximilians-Universit\"at M\"unchen}}
\centerline{{\em Theresienstr. 37, 80333 M\"unchen, Germany}}

\vspace{0.7truecm}

\begin{abstract}
For supersymmetric heterotic string compactifications on a Calabi-Yau
threefold $X$ endowed with a vector bundle $V$
the world-sheet superpotential $W$ is a sum of contributions
from isolated rational curves $\C$ in $X$; the individual contribution
is given by an exponential in the K\"ahler class of the curve times a prefactor
given essentially by the Pfaffian which depends on the moduli of $V$
and the complex structure moduli of $X$.
Solutions of $DW=0$ (or even of $DW=W=0$) can arise either by nontrivial
cancellations between the individual terms in the summation 
over all contributing curves or because 
each of these terms is zero already individually.
Concerning the latter case
conditions on the moduli making a single Pfaffian vanish 
(for special moduli values) have been investigated.
However, even if corresponding moduli - fulfilling these constraints - 
for the individual contribution of one curve are known
it is not at all clear whether {\em one} choice of moduli exists
which fulfills the corresponding constraints {\em for all contributing 
curves simultaneously}.
Clearly this will in general happen only if the conditions on the 
'individual zeroes' had already a conceptual origin 
which allows them to fit together consistently.
We show that this happens for a class of cases.
In the special 
case of spectral cover bundles we show that a relevant solution set has 
an interesting location in moduli space and is related to transitions
which change the generation number.
\end{abstract}

\newpage
\section{Introduction}

For a heterotic string compactification on a Calabi-Yau threefold $X$ 
endowed with a stable holomorphic vector bundle $V$ 
the world-sheet instanton superpotential $W$ 
is a sum of contributions
from isolated rational curves $\C$ (assumed to be smooth) in $X$
\beqa
W&=&\sum_{\C} \, W_{\C}
\eeqa
(we call such curves 'instanton curves').
The individual contribution is given by
\beqa
\label{individual Pfaffian}
W_{\C}&=&\frac{Pfaff_{\C}(v,z)}{D_{\C}(z)^2}\; e^{i C\cdot J}
\eeqa
where the Pfaffian depends on the vector bundle moduli of $V$ and the
complex structure moduli of $X$ (collectively denoted by $v$ and $z$, 
respectively), the denominator $D$ depends on $z$, and the exponential
term gives the dependence on the K\"ahler class $J=k_i J_i$ (we denote
the Kahler moduli collectively by $k$ and the homology class 
of $\C$ by\footnote{this is the notation in this section; 
in the rest of the paper $C$ denotes various spectral cover objects} 
$C$ where $C\cdot J=\int_{\C} J$). Actually 
$Pfaff_{\C}(v,z)=Pfaff(\bar{\partial}_{V(-1)}(\C))$
and $D_{\C}(z)=det \; \bar{\partial}_{\cO(-1)}(\C)$ times the constant 
$det'\; \bar{\partial}_{\cO}$ (here one considers twists by $\cO(-1)$
on the rational curve $\C$).\\
\indent
As the K\"ahler factor depends only on the homology class $C=[\C]$ of $\C$
one can write
\beqa
\label{Kahler split off}
W&=&\sum_C \Big( \sum_{[\C]=C} 
\frac{Pfaff_{\C}(v,z)}{D_{\C}(z)^2}\Big) \; e^{i C\cdot J}
\; =:\; \sum_C P_C \; e^{i C\cdot J}
\eeqa
\underline{\em remarks:} 1) Actually, because of the phase-factor subtlety 
[\ref{W}], the summation in (\ref{Kahler split off})
is conceptually more difficult; in 
$J=B+i\tilde{J}$ the $B$-field contribution has strictly speaking
to be considered together with the Pfaffian factor. 
Connected with this is that the naive
moduli space factorisation $\M \x \M_K=\M_V \x \M_{cx}\x \M_K$ does not hold:
clearly because of the holomorphy and stability condition for $V$
its moduli space $\M_V$ is actually fibered over $\M_{cx}\x \M_K$; but 
the phase-factor subtlety for the $B$-field is related to the anomaly
cancellation equation which connects its field strength $H$ to the quantity
$tr F\wedge F$ which depends in turn on $\M_V$. 
We will be able to ignore these issues in this paper.
We will write the naive product, and, when speaking of the full
$\M_{cx}$ or $\M_K$ as possible moduli sets for $v\in \M_V$ in
a solution, we mean the regions where $V_v$ is holomorphic or stable.\\
\indent
2) Another subtlety is the fact that the existence of holomorphic 
rational curves $\C$ depends on the complex structure moduli of $X$;
however, if $\C$ is isolated {\em and} even already rigid infinitesimally
(actually even {\em this} is necessary for contribution) 
then the normal bundle of $\C$ is $\cO(-1)\oplus \cO(-1)$
and $\C$ deforms with $X$ in all directions in $\M_{cx}(X)$; 
so such contributing curves do not arise newly by specialisation.\\
\indent
To place the following considerations in their proper context 
we distinguish first a number of different 
but closely related questions pertaining to $W$ and the $W_{\C}$.

\subsection{Direct investigation of the full superpotential $W$}

Here and in the following subsection we will distinguish in each case
considerations from the point of view 
of conformal invariance of the sigma model,
related to universal (in the moduli) vanishing results, from
considerations under the guiding philosophy to use actually a superpotential
(which does {\em not} vanish {\em identically}) to fix the moduli.

\subsubsection{The universal case}

The most immediate question, and the one which was originally first
investigated, when the world-sheet point of view prevailed, was naturally
the following (the term 'universal' refers to $W$ being {\em identically}
zero, independent of any special choice of moduli values):

\noindent
\underline{\em {\bf Question W(universal)}} \\
For which pairs $(X, V)$ is $W\equiv 0$ ?

If this holds for a pair $(X, V)$ then one must actually have a
seemingly even stronger result (but which is really equivalent 
as being enforced by $W\equiv 0$): as $W=0$ should hold independently
of any choice of specific values for the Kahler moduli, terms with 
Pfaffians for curves in different homology classes cannot cancel against
each other; that is, each Kahler prefactor $P_C$ has to vanish individually.
This gives the equivalent

\noindent
\underline{\em {\bf Question W(universal)$^{\prime}$}} \\
For which pairs $(X, V)$ is $P_C= \sum_{[\C]=C} 
\frac{Pfaff_{\C}(v,z)}{D_{\C}(z)^2}\equiv 0$ for all homology classes $C$ ?

Here one has the remarkable result 
[\ref{BW}] (assuming $c_2(V)=c_2(X)$)
that this holds
via a nontrivial cancellation between nonzero terms for example 
for hypersurfaces or complete intersections $X$ in an ambient 
(weighted) projective space $Y$,
with $V$ arising as restriction from a bundle on $Y$
(for a similar simpler case cf.~[\ref{SW}]).
Other cases, where this holds not from a nontrivial cancellation
between nonzero terms, but because of vanishing of each term individually,
will be mentioned below (cf.~Question Pfaff$_{total}$(universal)).

\subsubsection{The special case}

This first question above, the search for $W\equiv 0$, 
reflects directly the stringy or world-sheet point of view
and was much discussed in the late 80's. In the point of view which 
looks on an effective low energy action in four dimensions, which prevailed
in the mid-90's and later, issues like moduli stabilisation from such 
superpotentials were the prominent focus. Here the question is this:

\noindent
\underline{\em {\bf Question W(special)}}\\
For which points $(t,k)\in \M \x \M_K$ is $W(t, k)=0$ ?

A refinement of this question, relevant for solutions of the system 
$DW=W=0$, is:

\noindent
\underline{\em {\bf Question W(special)$^{{\bf \geq 2}}$}}\\ 
For which points $t\in \M$ is $W(t, k)=0$ to
second or higher order ?

Properly speaking, of course, instead of asking first where $W$ vanishes and 
then where it does so even to higher order, 
the first physical question (which is here that of moduli stabilisation
by demanding the supersymmetry conditions) amounts to where
in moduli space $DW(t,k)=0$ is solved; 
only then one evaluates the negative or zero
tree-level cosmological constant $V=-3|W|^2 e^K$ 
and may ask for Minkowski vacua which have also $W(t, k)=0$
(which reduces then, when posed as an additional condition, 
the covariant derivatives to ordinary ones, 
and leads one to pose the question above).

\subsection{\label{Pfaffian subsection}
Investigation of $W$ from the perspective of its individual
contributing summands from isolated rational curves $\C$}

The questions above represent 
one line of reasoning where one takes into consideration {\em from the outset}
the {\em total} world-sheet superpotential $W$,
with all individual contributions 
summed up. But as $W$ is a rather complex quantity
one also investigates, in a second line of investigation, 
first its individual contributions from the appropriate rational curves $\C$. 

\subsubsection{The universal case}

This leads to the following question (specific in $\C$, but universal
in the {\em identical} vanishing):\\
\noindent
\underline{\em {\bf Question Pfaff$_{\C}$(universal)}}\\ 
For which pairs $(X, V)$ 
and isolated (smooth) rational curves $\C$ is $Pfaff_{\C}\equiv 0$ ?\\
\indent
Here one has the criterion\footnote{the analysis is done for $SO(32)$ bundles
or $E_8\x E_8$ bundles with structure group in $SO(16)\x SO(16)$} 
[\ref{W}]
that this happens just if $V$, which restricts
to $\C$ as a direct sum of line bundles, is nontrivial, 
i.e.~$\neq \oplus \cO_{\C}(0)$.\\
\indent
In view of our goal to understand the {\em full} $W$
one is led naturally to the question:

\noindent
\underline{\em {\bf Question Pfaff$_{{\bf total}}$(universal)}}\\ 
For which pairs $(X, V)$ 
is $Pfaff_{\C}\equiv 0$ for all isolated (smooth) rational curves $\C$ ?

One case is the tangent bundle $V=TX$ 
as $V|_{\C}=\cO(2)\oplus \cO(a)\oplus \cO(-a-2)$ is always nontrivial.
For $X$ the generic quintic in ${\bf P^4}$
a stable deformation $V$ of $TX\oplus \cO$ 
exists having the required property for all lines [\ref{H}].
For yet another case cf.~[\ref{D}].

\subsubsection{\label{Pfaff special}The special case}

Now let us change again the perspective from results holding {\em universally
in the moduli} to the philosophy where {\em special moduli} 
are chosen from the conditions imposed.
So one now asks, in case that $Pfaff_{\C}$ 
does not vanish identically, again the specified question:

\noindent
\underline{\em {\bf Question Pfaff$_{{\bf \C}}$(special)}}\\
For which points $t\in \M$ is $Pfaff_{\C}(t)=0$ ?

Although one has the mentioned criterion of nontriviality of $V|_{\C}$,
to detect, whether this actually holds or not in dependence on the moduli, 
is still a nontrivial task. 

Again of interest, for finding supersymmetric solutions (which are in 
this approach necessarily Minkowski vacua), is also the refinement:

\noindent
\underline{\em {\bf Question Pfaff$_{{\bf \C}}$(special)$^{{\bf \geq 2}}$}}\\ 
For which points $t\in \M$ is $Pfaff_{\C}(t)=0$  to
second or higher order ?

Some examples, concerning the last two questions, 
were computed in a computer calculation
in [\ref{BDO}] and understood more conceptually in [\ref{C1}, \ref{C2}].

As in the universal case, 
in view of the fact that the total contribution in $W$ is a sum over all
contributions from the individual curves $\C$, the following question
is now of importance and {\em the one we focus on in the present paper}:

\noindent
\fbox{\em {\bf Question Pfaff$_{{\bf total}}$(special)}}\\ 
For which points $t\in \M$ is $Pfaff_{\C}(t)=0$ 
for all instanton curves $\C$ simultaneously ?

To use this approach to find supersymmetric (Minkowski) vacua one needs
control also over the derivatives and so one is 
{\em led again to the refinement}

\noindent
\fbox{\em {\bf Question 
Pfaff$_{{\bf total}}$(special)$^{{\bf \geq 2}}$}}\\ 
For which points $t\in \M$ is $Pfaff_{\C}(t)=0$ to second or higher order 
for all instanton curves $\C$ simultaneously ?

By posing theses questions we have replaced the original
system of conditions 
\beqa
\p_{v_l}W\; =\; \p_{z_m}W\; =\; \p_{k_j}W\; =\; W\; =\; 0, \;\;\;\;\;\;\;\;
\forall l, \forall m, \forall j
\eeqa 
for the total superpotential 
by the system of equations pertaining to the instanton 
contributions $Pfaff_{\C_i}$ from all the individual instanton curves $\C_i$
\beqa
\label{new system}
\p_{v_l}Pfaff_{\C_i}\; =\; \p_{z_m}Pfaff_{\C_i}\; =Pfaff_{\C_i}\; =\; 0 
\;\;\;\;\;\;\;\;
\forall l, \forall m \; \mbox{and} \; \forall i
\eeqa
(with $v_k, z_j$ and $k_m$ local coordinates 
on $\M_V, \M_{cx}$ and $\M_K$, resp.).
We will collect some general remarks about the relation of these systems
in sect.~\ref{Relation subsection} 
and especially in sect.~\ref{preliminary remarks} below
where we treat the moduli freedom in $\M_{cx}$ and $\M_K$
for solutions of (\ref{new system}) in detail.

\subsection{\label{Relation subsection}
Relation of these two lines of consideration}

Clearly, answers to the Question Pfaff$_{total}$(universal) 
represent also particular answers - though one might say the trivial ones 
(no cancellation between nonzero summands) - to the
Question W(universal); similarly answers to the 
Question Pfaff$_{total}$(special)
represent also particular answers - though, again, 
one might say the trivial ones (no cancellation) - to the
Question W(special) above 
(and similarly if one takes into account higher orders).

These assessments reflect the perspective looking from cases with an actual
cancellation between summands. 
However one can see the situation also from the 
following angle: let us assume we have investigated some individual Pfaffians
$Pfaff_{\C_i}(v,z)$ and found, for each $i$ individually, 
some solution sets, i.e.~$(v,z)\in \cS_i\subset \M  \Lra  Pfaff_{\C_i}(v,z)=0$
(the issue of higher multiplicities of the zeroes 
may be included in the discussion).

Then it is a priori quite surprising that a moduli choice
which makes the Pfaffian for $\C_1$ vanish should also do so for $\C_2$,
let alone for all $\C_i$. Although it is true that the question of vanishing
for $Pfaff_{\C_i}$ usually does not see all the moduli of $V$ (as some sort
of restriction to the part of the set-up relevant to $\C_i$ is involved; 
cf.~below) and therefore even a sharp (narrow) condition on the (restricted set
of) moduli relevant to $\C_i$ enlarges to a more relaxed condition 
when considered for the full set of moduli of $V$, it nevertheless 
should usually lead to a contradictory system of conditions when posed for all
$\C_i$ at once. 

So, from this perspective, the possibility of {\em getting zeroes of 
$W$ from getting individual zeroes for all the $Pfaff_{\C_i}$} 
(of suitable order)
looks almost
as mysterious as a potential cancellation of individual nonzero summands:
{\em here again some sort of hidden cooperation must happen to solve this now 
(usually) highly overdetermined system} (cf.~sect.~\ref{dim count}).

\subsection{Plan of the paper}

After some preliminary remarks in {\em sect}.~\ref{preliminary remarks}
we start our investigation in {\em sect}.~\ref{first try} in a set-up where 
results about the vanishing behaviour of individual Pfaffians were found:
the case of spectral cover bundles
over an elliptic Calabi-Yau space $\pi:X\ra B$.
Here we will be able to achieve a definitive answer to our main question(s)
mentioned in subsection \ref{Pfaff special}. However, 
our result will suffer from a serious drawback: in this approach
one has control over the vanishing behaviour of Pfaffians only for {\em base}
instanton curves $\C_i\subset B$.

Therefore we make a second start in {\em sect}.~\ref{new class}
where we define a new class of bundles with the leading idea
that the role played by the base $B$ in the spectral construction
is now played by the full (arbitrary) Calabi-Yau space $X$ itself.
For these bundles we succeed in giving a positive answer 
now for {\em all} instanton curves. In {\em sect}.~\ref{Conclusions}
we conclude.

\section{\label{preliminary remarks}Some preliminary remarks}

\resetcounter

After having noted already in sect.~\ref{Relation subsection}
that our general system of conditions
(where we use the notation $C_i=[\C_i]$ for the homology class, $D_i:=D_{\C_i}$
and $k\cdot C_i=\sum k_j n_j^{(i)}$)
\beqa
\label{original 1}
W&=&\sum_{\C_i}Pfaff_{\C_i} \; D_i^{-2}\; e^{i \, k\cdot C_i}\; = \; 0 \\
\p_{k_j}W&=&\sum_{\C_i} Pfaff_{\C_i}\; D_i^{-2} \; i\, n_j^{(i)} 
\; e^{i \, k\cdot C_i}\; 
= \; 0 \\
\p_{z_m}W&=&\sum_{\C_i} 
\Big[ (\p_{z_m}\, Pfaff_{\C_i}) \; D_i^{-2}\, 
+\, Pfaff_{\C_i}\; \p_{z_m}\, D_i^{-2}\Big]\; e^{i \, k\cdot C_i}
\; = \; 0 \\
\label{original 4}
\p_{v_l}W&=& \sum_{\C_i} (\p_{v_l}\, Pfaff_{\C_i})\; D_i^{-2}\; 
e^{i \, k\cdot C_i}\; = \; 0 
\eeqa
(for brevity of notation we suppressed the specific solution points 
in $\M_V\x \M_{cx}\x \M_K$ where all the expressions are actually evaluated
to give zero; similarly below)
admits as particular solutions the solutions of the other system
(called the 'special solutions' below)
\beqa
\label{specialised 1}
Pfaff_{\C_i}&=&0\\
\p_{z_m}\, Pfaff_{\C_i}&=&0\\
\label{specialised 3}
\p_{v_l}\, Pfaff_{\C_i}&=&0
\eeqa
we want to collect in this short section some (formal) notes on the possible
freedom in $\M_{cx}\x \M_K$ for solutions characterising (or not, cf.~below) 
such more special solutions.

\subsection{Remarks on the moduli freedom in $\M_{cx}\x\M_K$ for our 
special solutions}

We begin with the 'reduction' from $W$ to the 
$P_C=\sum_{[\C_i]=C}Pfaff_{\C_i}\, D_i^{-2}$; the possibility of a second 
'reduction', from the $P_C$ to the $Pfaff_{\C_i}$, will be discussed 
in sect.\ref{Resume} below. 

\noindent
\un{\em Remarks:}
a) Note that for 'special solutions' (which fix the moduli 
$(v,z)$ to lie in a certain region $S_{\M}$ in $\M=\M_V\x \M_{cx}$) 
the Kahler moduli remain completely flat directions.
Actually this will be the case 
for any solution which makes all the individual $P_C$ vanish 
(for certain values in $\M$). 
One has {\em also the converse}, i.e.~the {\em precise} association
\beqa
\un{\mbox{Solutions}}
\;\;\;\;\;\;\;\;\;\;\;\;\;\;\;\;\;\;\;\;\;\;\;\;\;\;\;\;\;\;\;\;\;
\;\;\;\;\;\;\;\;\;\;\;\;\;\;\;\;\;\;\;\;\;\;\;\;\;\;\;\;\;\;\;\;\;\;\;
\un{\mbox{Solutions}}\;\;\;\;\;\;\;\;\;\;\;\;\;\;\;\;\;\;\;\;\;\;\;\;\;\;\;\;
\;\;\;\;\;\;\;\;\;\;\;\;\;\;\;\;\;\;\;\;\;\;\;\;\;\;\;\;\;\;\;\;\;\;\;\;\;\;\;
\;\;\;\;\;\;\;\;\;\;\;\;\;\;\;\;\;\;
\nonumber\\
\mbox{with} \;\; \left\{ \begin{array}{ll}
\mbox{$\;\;(v,z)$ in a certain}\\
\mbox{region in $\M_V\x \M_{cx}$}
\end{array} \right\}
\;\;\;\;\;\;\;\;\;
\stackrel{=}{\longleftrightarrow }
\;\;\;\;\;\;
\mbox{where for }\;\left\{ \begin{array}{ll}
\mbox{$\;\;(v,z)$ in a certain}\\
\mbox{region in $\M_V\x \M_{cx}$}
\end{array} \right\}\;\;\;\;\;\;\;\;\;\;\;\;\;\;\;\;\;\;\;\;\;\;\;\;\;\;\;
\;\;\;\;\;\;\;\;\;\;\;\;\;\;\;\;\;\;\;\;\;\;\;\;\;\;\;\;\;\;\;\;\;
\;\;\;\;\;\;\;\;\;\;\;\;\nonumber\\
\mbox{times}\;\; \mbox{{\bf \{}flat K\"ahler directions{\bf \}}} 
\;\;\;\;\;\;\;\;\;\;\;\;\;\;\;\;\;\;\;\;\;\; \;\;\;\;\;\;
\mbox{one has {\bf \{} $P_C=\p_{z_m}P_C=\p_{v_l}P_C=0$, $\forall$ $C${\bf \}}}
\;\;\;\;\;\;\;\;\;\;\;\;\;\;\;\;\;\;\;\;\;\;\;\;\;\;\;\;\;\;\;\;
\;\;\;\;\;\;\;\;\;\;\;\;\;\;\;\;\;\;\;\;\;\;\;\;\;\;\;\;\;\;\;\;\nonumber
\eeqa

b) Recall that
the conditions $\p_i W=0=W$ for a supersymmetric Minkowski vacuum
are $h_m+1$ conditions for $h_m$ variables where $h_m=h_K+h_{cx}+h_V$
is the number of moduli, which comprises $h_K=h^{1,1}(X)$ K\"ahler moduli, 
$k_{cx}=h^{2,1}(X)$ complex structure moduli and $h_V=h^1(X, End\, V)$ 
vector bundle moduli. So, without the Minkowski condition this system of
$h_m$ equations (then of course with the covariant derivatives $D_i W=0$)
for $h_m$ complex variables should have generically a solution.
That a corresponding supersymmetric vacuum should be a Minkowski vacuum
makes the question nontrivial.

If, however, one of the equations $\p_i W=0$ would {\em not} pose effectively
a condition because one has $\p_i W \equiv 0$, i.e.~the derivative 
vanishes {\em identically} 
(the modulus labeled by $i$ being a {\em flat direction}), 
then one would be back to the generic counting 
argument, now even including the additional Minkowski condition $W=0$.
So one has the following

\noindent
\underline{\em {\bf Fact}} \\
If the moduli space of a pair $(X,V)$ contains a flat direction 
one will generically have a supersymmetric Minkowski vacuum.

Now, in our context the following distinction becomes important
which will be related to the existence of such flat directions in the
K\"ahler moduli part of $\M\x \M_K$.

\noindent
\underline{\em {\bf Definition 
'Generically K\"ahler-determined Bundles'}} \\
A pair $(X,V)$ is called 'generically K\"ahler-determined'
if the real span (of the homology classes) of its 
contributing ($P_C\not \equiv 0$)
instanton curves generates $H_2(X, {\bf R})$.

\noindent
Note that one is speaking here of contributing 
{\em classes} $C$ of instanton curves $\C$.
If $[\C]=\sum_{j=1}^{h_K}n_j C_j$ is a decomposition of an instanton
curve in a homology basis ($n_j\in {\bf Z}$; 
the $C_j$ are in general not classes of instanton curves) 
and $J=\sum_{m=1}^{h_K}k_m J_m$ a dual decomposition
of a general K\"ahler class ($k_m$ are the K\"ahler moduli) 
such that $\C \cdot J=\sum n_j k_j$ one has
\beqa
\p_{k_m}W&=&\sum P_C \, in_m \, e^{i\C\cdot J}
\eeqa
If $V$ over $X$ is not generically K\"ahler-determined 
the span of contributing instanton
classes will not contain, say, $C_{h_K}$:
all contributing $C$ will have $n_{h_K}=0$ so that $\p_{k_{h_K}}W\equiv 0$.

\noindent
\underline{\em {\bf Fact}} \\
A bundle $V$ over $X$ which is not generically K\"ahler-determined 
has a flat direction.

So bundles which are not generically K\"ahler-determined will generically 
have a supersymmetric Minkowski vacuum. But 
usually interesting bundles will be generically K\"ahler-determined 
such that, {\em despite} the generic counting argument 
against a supersymmetric Minkowski vacuum,
the question arises for suitable {\em special} moduli choices providing such a
vacuum (cf.~also sect.~\ref{base relevance}).
As remark a) above shows the moduli choices in {\em our solutions}
will nevertheless lead to completely flat K\"ahler directions.

c) After the K\"ahler moduli, which we considered more closely 
in remarks a) and b), let us also have a closer look 
on the complex structure moduli.

We start with a general remark.
One may sometimes encounter special situations where each Pfaffian 
$Pfaff_{\C_i}(v,z)$ turns out to be independent of any complex structure
modulus $z_m$. 
We want to note here that this {\em does not mean} that the complex structure
moduli are flat directions for $W$: there is still the denominator $D(z)^2$
in (\ref{individual Pfaffian});
furthermore it does also not mean that in such a case 
(of $\p_{z_m}Pfaff_{\C_i}(v,z)\equiv 0$ everywhere 
$\forall i, \forall m$)
the solution set in $\M=\M_V \x \M_{cx}$ has the structure of a certain
region in $\M_V$ times the full $\M_{cx}$ factor.\footnote{'Solutions' include
here always the vanishing at least to second order, i.e.~we are speaking about
the system $\p_{k_j} W=\p_{z_m} W=\p_{v_l} W=W=0$.} 
But we will be anyway generally {\em not} in such a situation. 

Before we go on let us now remark that the special class
of solutions we will investigate in this paper {\em is indeed} actually
a class of solutions (cf.~subset.~\ref{Relation subsection} below): 
for if one has $Pfaff_{\C_i}(v,z)=0$, $\forall i$,
for $(v,z)$ in some solutions set $S\subset \M=\M_V\x\M_{cx}$
and has that this vanishing happens even to second or higher order
(such that also $\p_{v_l}Pfaff_{\C_i}(v,z)=0, \forall i, \forall l$ and 
$\p_{z_m}Pfaff_{\C_i}(v,z)=0, \forall i, \forall m$ hold in $S$)
then indeed one will have $S\x\M_K$ 
(for the final factor cf.~remark a) above)
as a solution set for our main system
of conditions $\p_{v_l}W=\p_{z_m}W=\p_{k_j}W=W=0, 
\forall l, \forall m, \forall j$.

Connecting now the considerations of the last two paragraphs one finds
that {\em under the assumption that the Pfaffians
are independent of the complex structure moduli} 
one has\footnote{Here and in remark a) above when we speak of {\em solutions}
with some flat directions - K\"ahler  moduli or complex structure moduli -
we mean of course only that the {\em solution set} contains a full factor
$\M_K$ or $\M_{cx}$ (and not that $W$ generally would be independent of 
the corresponding moduli).}
\beqa
\un{\mbox{Solutions}}
\;\;\;\;\;\;\;\;\;\;\;\;\;\;\;\;\;\;\;\;\;\;\;\;\;\;\;\;\;\;\;\;\;
\;\;\;\;\;\;\;\;\;\;\;\;\;\;\;\;\;\;\;\;\;\;\;\;\;\;\;\;\;\;\;\;\;\;\;
\un{\mbox{Solutions}}\;\;\;\;\;\;\;\;\;\;\;\;\;\;\;\;\;\;\;\;\;\;\;\;\;\;\;\;
\;\;\;\;\;\;\;\;\;\;\;\;\;\;\;\;\;\;\;\;\;\;\;\;\;\;\;\;\;\;\;\;\;\;\;\;\;\;\;
\;\;\;\;\;\;\;\;\;\;\;\;\;\;\;\;\;\;\;\;\;\;
\nonumber\\
\mbox{with} \;\; \left\{ \begin{array}{ll}
\mbox{$v$ in a certain}\\
\mbox{region in $\M_V$}
\end{array} \right\}
\;\;\;\;\;\;\;\;\;\;\;\;\;\;
\stackrel{\stackrel{\supset}{ \neq}}{\longleftrightarrow}
\;\;\;\;\;\;\;\;\;\;
\mbox{where for }\;\left\{ \begin{array}{ll}
\mbox{$(v,z)$ in a certain}\\
\mbox{region in $\M_V\x\M_{cx}$}
\end{array} \right\}\;\;\;\;\;\;\;\;\;\;\;\;\;\;\;\;\;\;\;\;\;\;\;\;\;\;\;\;\;
\;\;\;\;\;\;\;\;\;\;\;\;\;\;\;\;\;\;\;\;\;\;\;\;\;\;\;\;\;\;\;\;\;\;
\;\;\;\;\;\;\;\;\;\nonumber\\
\mbox{times}\;\; \mbox{{\bf \{}flat cx.~str.~directions{\bf \}}} 
\;\;\;\;\;\;\;\;\;\;\;\;\;\;\;\;\;\;\;\;\;\;\;\;\;\;
\mbox{one has}\;\left\{ \begin{array}{ll}
\mbox{$\;\;\; Pfaff_{\C_i}= 0$ for all $i$}\\
\mbox{to second or higher order}
\end{array} \right\}
\;\;\;\;\;\;\;\;\;\;\;\;\;\;\;\;\;\;\;\;\;\;\;\;\;\;\;\;\;\;\;\;\;\;\;\;
\;\;\;\;\;\;\;\;\;\;\;\;\;\;\;\;\;\;\;\;\;\;\;\;\;\;\;\;\;\;\;\;\;\;\nonumber
\eeqa
\indent
However, the assumption $\p_{z_m}Pfaff_{\C_i}\equiv 0$ is rather strong.
One point of {\em our specific explicit solution 
set}\footnote{which belongs to the wider, though still rather special class 
of solutions where one has
vanishing - to second or higher order - for each 
Pfaffian individually (for certain moduli); cf.~also sect.~\ref{Resume}} 
$\Si$ will be that it has flat 
$z_j$ directions (cf.~also sect.~\ref{Resume})
{\em even without making such an extreme assumption}:
more precisely the solution set
which we will find for each individual Pfaffian will have such a structure
$\Si_i \x \M_{cx}$ (concerning notational accuracy here cf.~later the
footn.~\ref{pullback footnote});
this will be 'globalised', to make all Pfaffians vanish
simultaneously, by going to the solution set $\Si \x \M_{cx}$ in $\M$.

\subsection{\label{Resume}R\'{e}sum\'{e}}

Above, in remark a) of sect.~\ref{preliminary remarks}, we showed how,
to be in a case where one has
the freedom to have all possible K\"ahler moduli of $\M_K$ 
as part of the solution set of our original system
(\ref{original 1}) - (\ref{original 4}) 
(2.~order refers to the value and all first derivatives)
\beqa
W&=&\sum_C P_C \, e^{i\, k\cdot C}\; =\; 0 
\;\;\;\;\;\;\;\; \mbox{to at least 2. order}
\eeqa
(like it happens for our specialised system 
(\ref{specialised 1}) - (\ref{specialised 3})),
characterises just those solutions 
for which already the individual $P_C$ vanish (to at least 2.~order).
This is possible as the 
functions $e^{i\, k\cdot C}$ for the different
homology classes are sufficiently 'separating' such that, having $W=0$
with the K\"ahler moduli $k_j$ in $k\cdot C=\sum k_j n_j^{(C)}$ 
nevertheless running completely unconstrained, can hold only if already
the $P_C$ are zero individually 
(for some $(v,z)\in S_{\M}\subset \M_V\x\M_{cx}$;
similarly for the derivatives).

\noindent
One now would like to characterise in a second 
'reduction step'\footnote{One highly exceptional case where some reduction 
{\em can} be effectuated
is the case when all contributing instanton curves are {\em distinguished}
in the sense that they are the only contributing curve in their class $C$:
then one finds indeed (from $D_i^{-2}\neq 0$) that our special solutions
{\em are} just those fulfilling 'only' (\ref{P_C reduction}).} 
our special solutions of $Pfaff_{\C_i}=0$ (to at least 2.~order)
among the already specialised class of solutions with 
\beqa
\label{P_C reduction}
P_C&=&\sum_{[\C_i]=C}Pfaff_{\C_i}\; D_i^{-2}\; =\; 0
\;\;\;\;\;\;\;\; \mbox{to at least 2. order}
\eeqa
The naive expectation would be that now the $D_i(z)$ play the 
role of the exponential factors and one is reduced to solutions
where also the complex structure moduli run free.

However, two facts get in the way of this potential second reduction:
the expressions $D_i(z)$ in the complex structure moduli
are not given as explicitly as the exponentials in the K\"ahler moduli;
secondly, the pfaffians $Pfaff_{\C_i}(v,z)$ itself depend also on 
the complex structure moduli $z_m$.
This has the consequence that there are various possibilities which 
inhibit in general the intended reduction: there may be the case, for example,
that the $Pfaff_{\C_i}$ (of a fixed class $C$) do not depend actually
on the $z_m$ and that the $D_i$ are independent of the concrete instanton curve
$\C_i$ of class $C$; then, easily, bundle moduli $v$ may exist for which
$\sum_i Pfaff_{\C_i}(v)=0$ (to order $\geq 2$)
which leads to $P_C=0$ (to order $\geq 2$) with the $z_m$
being able to run freely (unconstrained) without that one would need to have 
$Pfaff_{\C_i}(v)=0$ (to order $\geq 2$) individually. 
Another (theoretical) 
possibility is that the $z$-dependence in $Pfaff_{\C_i}(v,z)$ cancels
that of $D_i(z)$ and one finds a $v$ as before.

\noindent
So the set of solutions with $P_C=0$ (to at least 2.~order)
already (or equivalently
solution set $S=S_{\M}\x\M_K$) has two distinct subsets, characterised
by $S=S_V\x \M_{cx}\x\M_K$ and $Pfaff_{\C_i}=0$, resp..
Our {\em concrete} solutions will lie in the intersection of these two sets.

\subsection{\label{dim count}Dimension counting}

Before closing this propaedeutic section with its preliminary, formal remarks
we want to state the dimension counting which shows how nontrivial in general
is the idea to solve for supersymmetric Minkowski vacua 
(moduli leading to such vacua we call just 'solutions')
by looking for critical
points (including zero value) of all the individual Pfaffians of the
instanton curves which contribute to the sum which builds the superpotential
$W$.

Now, the ordinary approach or direct approach, involving just $W$ and
its derivatives, leads to
\beqa
\un{{\bf \mbox{General approach}}}\;\;\;\;\;\;\;\;\;\;\;
h_m+1 \;\;\;\; \mbox{conditions}\;\;\;\;\;\;\;\;\;\; \mbox{for} 
\;\;\;\;\;\;\;\;\;\;\;\;\;\; h_m \;\;\;\; \mbox{moduli}\;\;\;\;\;\;\;\;\;
\eeqa
where $h_m=h_K+h_{cx}+h_V$ counts the number of all moduli,
comprising the K\"ahler, complex structure and bundle moduli.
Here, as remarked earlier on, the additional condition comes from $W=0$,
thus securing that the supersymmetric vacuum is Minkowski.
So the system is only mildly overdetermined

By contrast, in our approach for the 'special solutions',
involving all $N$ Pfaffians $Pfaff_{\C_i}$ and all of their first order 
derivatives, one has
\beqa
\un{{\bf \mbox{'Special solutions'}}}\;\;\;\;
N\cdot (h+1)\;\; \mbox{conditions}\;\;
(\mbox{with} \; N=\sharp\{\C_i\}) \;\; \mbox{for} 
\;\; h \;\;\;\; \mbox{moduli}\;\;\;\;\;
\eeqa 
where $h=h_{cx}+h_V$ counts the number of moduli on which a Pfaffian depends,
thus comprising only the complex structure and bundle moduli.
Thus, as was to be expected, the system for the 'special solutions' 
shows a massive overdetermination.
It seems to be a miracle how such a procedure should lead to solutions at all;
and certainly the existence of such special solutions should, 
as remarked earlier on, only be possible if the conditions coming from
the $N$ different instanton curves somehow 'fit together nicely'.

Now, what actually happens in our {\em concrete} examples for 
the occurrence of special solutions in certain set-ups, is the following.
First a remark concerning our procedure.
We treat in the following two main sections \ref{first try} and 
\ref{new class} actually two set-ups: $SU(n)$ spectral cover bundles on a
Calabi-Yau space $X$ which is elliptically fibered over a base $B$ 
and then certain bundles 
defined in an analogy to the spectral construction for a general $X$.
The first set-up will give us control only over instanton curves in $B$;
but as the second construction in sect.~\ref{new class} proceeds in 
a certain analogy to the spectral construction which is already well known
we describe here the outcome for the first case, the spectral bundles on
an elliptic $X$ of sect.~\ref{first try} (so one has to keep in mind the 
caveat that the results in this case are only rudimentary as the concern
actually only the part $W_B$ of $W$ build by contributions from base curves).

Now, two simplifications (described here in the set-up of sect.~\ref{first try})
occur in our {\em concrete} example of such special solutions
(despite our description both reductions occur at the same time). 
Consider first a single Pfaffian $Pfaff_{\C_i}(v,z)$.
It turns out, unexpectedly, that there is no loss of dimensions
connected with the demand that not only the value but also all first
derivatives $\p_{v_l}Pfaff_{\C_i}(v,z)$ and $\p_{z_m}Pfaff_{\C_i}(v,z)$
vanish: the additional vanishing to second order can be tuned topologically,
i.e.~by choosing discrete parameters in a suitable 
manner\footnote{cf.~part b) of Theorem $Pfaff_{all\, \C\subset B}$ below in 
sect.~\ref{concrete solution set}}. 
So in this concrete case one gets a reduction of conditions (which concern the 
counting of degrees of freedom of the continuous parameters):
\beqa
\un{{\bf \mbox{'Concrete special solutions'}}}\;\;\;\;
N\cdot (h+1)\;\; \mbox{conditions}\;\;\;\; \lra
\;\; N\cdot 1 \;\;\;\; \mbox{conditions}\;\;\;\;\;
\eeqa

Also a second simplification occurs:~the different conditions for all
the individual Pfaffians related to the respective instanton curves $\C_i$
fit together nicely to a set of conditions which are independent of the
respective $\C_i$ and can be posed for the global bundle $V$ once and for all.
Thereby the number $N$, which is 'dangerous' as it can be {\em quite large}
(examples show values in the thousands) and, more importantly,
{\em independent}\footnote{in particular there is a priori no reason 
that $N$ should be not greater than $h$} 
{\em of the number of bundle (and complex structure) moduli}, 
is reduced to a number which is 
related directly to bundle data 
(and independent of the 
number\footnote{which depends mainly just on the geometry of $X$; only 
through the issue which of the $\C_i$ actually are contributing for $V$
(i.e.~for which of them one has $Pfaff_{\C_i}\not\equiv 0$) it 
is related to $V$} 
$N$)
and can be shown not to be greater than 
the relevant number $h$ of moduli (for the notation cf.~sect.~\ref{first try}):
\beqa
\un{{\bf \mbox{'Concrete special solutions'}}}\;\;
N\cdot 1\;\; \mbox{conditions}\; \lra
\; (n-2)\frac{1}{2}\eta(\eta-nc_1) \;\; \mbox{conditions}\;\;\;\;
\eeqa

The mentioned final number of conditions (on continuous parameters) 
arises from the explicit description of the solution set
which specialises (via $n-2$ divisibility conditions)
the coefficients $a_j$ in the spectral cover equation
of the cover surface $C$ 
\beqa
\R  &=& \Big\{ t\in \M_{X}(C)\Big| \; 
a_{n} | a_{j}\;\; \mbox{for}\;\; j=2, \dots , n-1\Big\}
\eeqa
(this will be explained below).
It gives the codimension in the continuous part of the
bundle moduli space (for the concrete counting cf.~remark a) after 
Theorem $Pfaff_{all\, \C\subset B}$ in sect.~\ref{concrete solution set}; 
for our {\em concrete} special solutions the
complex structure moduli remain unconstrained; 
as for {\em all} special solutions
the K\"ahler moduli remain also unconstrained).

As emphasized, all of this concerns the set-up of sect.~\ref{first try}
with its rudimentary results (pertaining only to base instanton curves); 
a corresponding discussion, proceeding along analogous lines,
may now ensue for the proper case (pertaining to all instanton curves)
of our original problem in the set-up of sect.~\ref{new class}.

\section{\label{first try}
A first try: spectral cover bundles on elliptic Calabi-Yau spaces}

\resetcounter

To understand vanishing conditions for various Pfaffians 
it is helpful to discuss first in greater detail an explicit example.
For this reason we start in this section by treating the case of
a spectral cover bundle on an elliptically fibered Calabi-Yau space.
Here we will be able to formulate quite definitive conditions
for the vanishing of a single Pfaffian, in case it is related to
an instanton curve {\em in the base} of the elliptic fibration.
This will be then extended to a coherent treatment of the
conditions for {\em simultaneous} vanishing of all the Pfaffians related 
to instanton curves in the base. The significance of this restricted
success will be discussed below in sect.~\ref{base relevance} 
and \ref{location subsection}. We will go beyond this in sect.~\ref{new class}.

So let us assume in this section that $X$ 
admits an elliptic fibration $\pi: X\rightarrow B$ which has 
a section\footnote{thereby $B$ will be considered to lie in $X$; we
denote the co-/homology class of $B$ by $\sigma$, $F$ is the fibre; 
furthermore we denote $c_1(B)$ just by $c_1$ and usually will suppress
pullbacks like in $\pi^*\eta$ for $\eta\in H^{1,1}(B)$} 
$\sigma$.
Let us also make at first the (unreal) assumption
that all potentially contributing isolated (smooth)
rational curves $\C_i, i=1, \dots , p$ lie in $B$; we will discuss the 
real, more complicated situation in the next section.
Then $Pfaff_{\C_i}$ will depend on the moduli $v$ of $V$ only through
the restriction $V|_{\E_i}$ of $V$ to the elliptic surface 
$\E_i=\pi^{-1}(\C_i)$ over $\C_i$. Let us also assume that $V$ is an $SU(n)$ 
spectral cover bundle with spectral cover surface $C$ 
(of class $n\si + \pi^* \eta$), given by an equation
(for standard technical details of the spectral cover description
cf.~app.~\ref{spectral cover appendix})
\beqa
\label{C equation}
w=a_0+a_2 x + a_3 y + \dots + a_n x^{n/2}=0
\eeqa 
(for $n$ even, say; 
the coefficients $a_j$ are sections over $B$ of suitable line bundles); 
so, apart from a discrete choice of a twist parameter $\lambda$, the moduli
of $V$ are given by the motions of $C$ in $X$ 
(essentially, up to an overall scaling, they are
the coefficients of the sections $a_j$; here by coefficients we mean the
degrees of freedom, ordinary coefficients for $B={\bf F_k}$)
\beqa
\M_X(C)&=&{\bf P}H^0(X, \cO(C))
\eeqa
The geometric meaning is that the bundle $V$ decomposes over the 
generic elliptic fibre $F$ as a direct sum of line bundles $\cO_F(q_i-p)$
where the point $p$ is the zero point (where the section $\sigma$ meets $F$)
and the $q_i, i=1, \dots, n$ are some other points where $C$ meets $F$.
These point positions on each fibre are the continuous moduli of the problem
and are encoded by the (degrees of freedom of the)
coefficients $a_j$ of the equation (\ref{C equation}) for $C$. 

The other discrete degree of freedom mentioned above, 
the twist parameter $\lambda$, arises as follows: when reconstructing
the bundle $V$ from the fibrewise data it is still possible to twist
the construction by a line bundle $L$ on $C$ 
which is cohomologically fixed\footnote{and even completely fixed up to the
mentioned discrete choice if one assumes, as we do, that $H^{0,1}(C)=0$ (as
would follow, for example, for $C$ being an ample divisor 
in $X$ from $H^{0,1}(X)=0$)}
from our standing assumption $c_1(V)=0$ 
with the only remaining freedom in $c_1(L)$ being an element
$\ga\in ker \, \pi_{C*}: H^{1,1}(C)\ra H^{1,1}(B)$; but $\ga$ can generically 
be only of the form
\beqa
\label{gamma class}
\ga&=&\la \, \Lambda\; = \; \la \, \widetilde{\Lambda}|_C
\; = \; \la \, \Big(n\si - (\eta - n c_1)\Big)\Big|_C
\eeqa
One finds $\la$ to be a half-integral integer 
with precise integrality restrictions for the various cases ($n$ even or odd).
From here one can write $L$ explicitly as a line bundle $\cO_X(D)|_C$.

The moduli space $\M_V$ of connected component $\M_X(C)$ has dimension 
\beqa
dim \, \M_V&=&h^0(X, \cO_X(C))-1=\frac{1}{12}\Big(c_2(C)+c_1^2(C)\Big)C-1
=\frac{1}{12}\Big(c_2(X)C+2C^3\Big)-1\;\;\;\nonumber\\
\label{full moduli space dimension}
&=&n-1+\frac{n^3-n}{6}c_1^2+\frac{n}{2}\eta(\eta-nc_1)+\eta c_1
\eeqa
Here we used the assumption that $C$ is ample 
(usually assumed so that $h^{0,1}(C)=0$ and $L$ has no continuous moduli). 
Below in (\ref{dim M_V}) 
we write the dimension from the degrees of freedom 
provided by the coefficients $a_j$ in (\ref{C equation}).

\subsection{Instanton curves $\C_i$ in the base $B$ and the relevant moduli}

Then the restrictions see only the motions of $c_i=C\cap \E_i$ in $\E_i$,
given essentially by the restrictions $a_{j;i}:=a_j|_{\C_i}$ which are 
homogeneous polynomials on these ${\bf P^1}$'s.
Now, from $Pfaff_{\C_i}(v,z)=0$ one gets ('narrow') conditions on all of the
polynomial coefficients of the $a_{j;i}$ 
which translate back to conditions on the $a_j$ 
(the 'more relaxed' conditions mentioned above as for each $\C_i$
only some subset of the moduli $a_j$ is seen by the restriction $a_{j;i}$). 
Now the point is that it is not at all clear that posing all
these conditions on $a_j$ for $i=1, \dots , p$ simultaneously still allows
for a solution. To allow for a common solution
the conditions should better 'fit together nicely'. 
Clearly this can not be expected to happen in general; 
rather these conditions will usually turn out to be contradictory.

However, and this is the point of the present note, there are cases
where the individual conditions from the $\C_i$ lead to a compatible set
of conditions on the moduli of the global bundle $V$ because each of these
conditions did express already not an 'accidental' zero for $Pfaff_{\C_i}$
but did represent a 'conceptual' zero: the latter
fits into a global system of conditions in the sense that the individual
conditions for the $V|_{\E_i}$ were already parts (concretely restrictions)
of one global condition on the bundle $V$ over $X$.
This is the case we are going to consider.

\subsection{The problem of skew curves}

However, there is a problem. In the spectral cover scenario
one can get, as we are going to describe, some control over the Pfaffians
for base curves $\C$ (as always smooth isolated rational), i.e.
curves in the image of $i:B\ra X$ (embedding via section). 
There are, of course, many other curves in $X$ relevant as instantons 
besides these 'horizontal' curves. Here we use the obvious terminology
calling a curve of homology class $i_* \psi + k F$ 'horizontal' 
(or a base curve) for $k=0$ (these are just the curves in $B$), 
'vertical' for $\psi=0$ 
(these [reducible for $k > 1$] fibers are here elliptic and therefore
not interesting for us), and finally we call them lying 'skew' in the remaining
cases. There can, and will in general, exist smooth isolated rational curves
in $X$ which are not lying in the base but are skew. 
The behaviour of the bundle,
with respect to the triviality criterion concerning the Pfaffian, 
is out of control for such non-base curves (as one lacks the crucial relation
$V|_{\C_i}=\pi_* L|_{c_i}$ from the spectral approach which is
tied to the fibration structure {\em over} $\C_i$). 

\subsection{A way out through a moduli split ?}

If one would know, however, that those moduli $v_s$ 
of $V$ which control the Pfaffians for the skew
curves are distinct from the corresponding moduli $v_b$ for the base curves
one could still argue that the common solution 
set\footnote{this in our examples turns
out always to be a whole submanifold and not just points},
where the Pfaffians vanish (to second order at least) 
in the 'base-relevant' moduli, 
constitutes a partial moduli fixing: for if one would have 
a decomposition $W=W(v_b)+W(v_s)$ this would mean
that, when taking derivatives with respect to the $v_b$,
the contribution from the $v_s$ does not matter.
In [\ref{BO}] it was claimed indeed that, because generically the skew curves
avoid to hit the base curve\footnote{there a {\em single} base curve 
is considered 
but the argument extends to the case of finitely many $\C_i\subset B$}, 
the corresponding moduli sets, relevant for the Pfaffian contributions, 
are disjoint. 

As a minor caveat let us first remark that, of course, 
one would in principle (when stabilising moduli and just demanding
that $W=0$ in total for the stabilised values)
not have needed to ask for $W=0$ individually for the $W(v_b)$ part
(as one does, however, here in this approach implicitly when using a 
second or higher vanishing order of the $W(v_b)$ part at its stabilised moduli).
Doing so, one has then also to ask for $W(v_s)=0$ individually for the
$W(v_s)$ part when stabilising the $v_s$ moduli (if one searches for moduli
stabilised with $W=0$ for the total $W$). It might occur a case, however,
that a solution exists which makes all the individual derivatives vanish
but makes also only the total sum $W=W(v_b)+W(v_s)$ vanish and not the 
individual parts.

More fundamentally, however, it is not at all clear whether one should even
expect that the set of all bundle moduli decomposes into sectors of moduli, 
relevant for the respective individual instanton curves, 
as soon as the latter are disjoint (a connection is made in [\ref{BO}] to
the problem of fivebrane-transitions and their relevant moduli where 
a fivebrane curve (the compact part of the fivebrane) is understood
as a limiting 'small instanton' object for the bundle and then dissolved
into the bundle, enlarging its second Chern class correspondingly;
but this is not investigated at all for skew curves).

Actually, the argument in [\ref{BO}] served, however, a different purpose
which can be achieved in another way. There the emphasis was to show that, 
after finding nontrivial results for individual Pfaffians, the full $W$ 
is not nevertheless identically zero from a cancellation 
after summation over all curves. 
Concretely there was a single base instanton curve 
and one had to care that its nonzero contribution is not cancelled.
For this one does not need, however, the mentioned unreal moduli split
(cf.~the discussion before Question W(universal)$^{\prime}$ in the 
introduction). It is
enough to note that a hypothetical fact $W\equiv 0 $ from 
cancellation between nonzero summands nevertheless must mean that the
individual $P_C=\sum_{[\C]=C} \frac{Pfaff_{\C}(v,z)}{D_{\C}(z)^2}$ already
vanish identically (cf.~the discussion earlier on). But obviously a base curve
is homologically distinguished (by definition) from a skew curve.

\subsection{\label{concrete solution set}The moduli space and the solution set}

We now give the details concerning the actual solution sets 
in our set-up. We have
a spectral cover bundle with (\ref{C equation}) as equation
of the cover surface $C$; we denote by 
$a_{n;i}=a_n|_{\C_i}=\prod_{k=1}^{r_i-n} a_{n;i}^{(k)}$ the
decomposition into linear factors of the highest coefficient 
($r_i:=\eta\cdot \C_i$). 
Further we define, following [\ref{C2}], in the moduli space $\M_{\E_i}(c_i)$
two subloci: first the locus 
$\R_i$ where all resultants $R_{j;i}^{(k)}=Res(a_{j;i}, a_{n;i}^{(k)})$ 
for $j=2, \dots , n-1$ and $k=1, \dots, r_i-n$ vanish;
this locus of codimension $(n-2)(r_i-n)$ is the locus
where $a_{n;i}$ is a factor of all the $a_{j;i}$ for $j=2, \dots , n-1$ 
\beqa
\R_i \; = \; \Big\{ t\in \M_{\E_i}(c_i)\Big| \; \;
a_{n;i} | a_{j;i}\;\; \mbox{for}\;\; j=2, \dots , n-1\Big\}
\eeqa
Secondly the locus where $\cO_{c_i}(\Lambda|_{c_i})$ becomes trivial
(using the notation $\Lambda$ also as divisor)
\beqa
\Sigma_i& = & \Big\{ t\in \M_{\E_i}(c_i)\Big| \; \; 
\cO_{c_i}(\Lambda|_{c_i})\cong \cO_{c_i}\Big\}
\eeqa
This means that as the moduli change (concretely as one has different
coefficient sections $a_j$ in the defining equation (\ref{C equation}) of $C$)
one has different concrete divisors given by the surface $C$ in $X$,
and correspondingly also different $c_i$; and as such a $c_i$ varies it may
happen that the universal twist bundle\footnote{A remark concerning notation: 
here $\Lambda$ denotes the divisor whose cohomology class 
was used in (\ref{gamma class}); 
this and corresponding occasional usage of 
common notation for some divisors and their cohomology classes 
will not cause any confusion} 
$\cO_C(\Lambda)$ on $C$ 
becomes trivial on $c_i$ for certain moduli.
As we recall in the next theorem this can be controlled on the one hand
by an explicit condition (related to the locus $\R_i$) in the $a_j$
and on the other hand leads to an effectivity assertion for the divisor
of the line bundle on $c_i$ which gives $V|_{\C_i}$ when projected down 
to $\C_i$; this effectivity makes then the corresponding Pfaffian vanish
as will be recalled in Theorem $Pfaff_{single\, \C\subset B}$ below.

In a similar vein one defines corresponding notions pertaining to the 
whole of the global surface $B$ (not just to a sublocus given by an
instanton curve $\C_i$)
\beqa
\R  &:=& \Big\{ t\in \M_{X}(C)\Big| \; 
a_{n} | a_{j}\;\; \mbox{for}\;\; j=2, \dots , n-1\Big\}\\
\Si &:=&  \Big\{ t\in \M_{X}(C)\Big| \; \;
\cO_C(\Lambda) \cong \cO_C\Big\}\;\;\;
\eeqa
So in both cases - the situation over a single instanton curve $\C_i$ in $B$
and the situation over $B$ as whole (with {\em all} instanton curves 
$\C_i\subset B$ considered simultaneously) -
we have defined two subloci of the relevant moduli space:
the locus $\Si_i$ or $\Si$ has a conceptual description and is
directly relevant to the vanishing of the Pfaffian; the set $\R_i$
or $\R$ has a completely explicit description. 
$\R_i\subset \Si_i$ or $\R\subset \Si$ is easily seen and it would suffice
to state our theorems for $\R_i$ or $\R$; but it is useful to have the 
following (proven in app.~\ref{Characterisation})

\noindent
{\bf \un{Proposition}} One has the following equalities of loci 
in the respective moduli spaces\\
a) $\R_i=\Si_i$ in $\M_{\E_i}(c_i)$\\
b) $\R=\Si$ in $\M_X(C)$\\
\indent
After these preliminary remarks on the geometric significance of the 
various subloci in moduli space which we have defined let us state the
relevance of these loci to our main problem, that is to the question
of the vanishing loci (vanishing divisors), denoted $(Pfaff_{\C_i})$, of the
various Pfaffians over instanton curves. In this section we will have
to restrict us to instanton curves $\C_i$ {\em which lie in the base $B$}.
As above we will first give the result for an individual instanton curve $\C_i$,
where we describe a subset of $(Pfaff_{\C_i})$,
and afterwards the globalised version describing the corresponding
subset of $\bigcap_{\C_i\subset B}(Pfaff_{\C_i})$.\\
\indent
In an investigation [\ref{C2}] about the conceptual meaning of some explicit 
components (and more general subloci) of the 
vanishing locus given by the zero divisor $(Pfaff_{\C_i})$ 
of Pfaffians $Pfaff_{\C_i}$ in some concrete examples [\ref{BDO}], 
the following Theorem was shown [\ref{C2}]
(for convenience of the reader we recall the proof of part a) and c) 
of the theorem in app.~\ref{proof appendix}; 
note that the $SU(2)$ assertion of b) follows from
a) as then $\R_i=\M_{\E_i}(c_i)$ as only $a_{0;i}$ and $a_{2;i}$ exist;
for further interpretations and some caveats cf.~sect.~\ref{base relevance} 
and \ref{location subsection}):

\noindent
{\bf \underline{Theorem}}  \fbox{$Pfaff_{single\; \C \subset B}$}\\
a) The set $\Sigma_i$ is contained in the locus
where $Pfaff_{\C_i}$ does vanish 
\beqa
\label{vanishing locus}
\Sigma_i \; \subset \; (Pfaff_{\C_i})
\eeqa
b) one has $Pfaff\equiv 0$ for $n=2$  \\
c) for the multiplicity $k$ of such a zero of $Pfaff_{\C_i}$, 
i.e. for the vanishing order, one has the lower bound
(where $r_i=\eta \cdot \C_i$)
\beqa
\label{multiplicity}
k&\geq & \left\{ \begin{array}{ll}
1+ \frac{n}{4}(r_i-\frac{n}{2})
\;\;\;\;\;\;\;\;\;\;\;\;\;\;\;\;\;\;\;\; 
\mbox{for} \;\; n\, \equiv \, 0 \, (2)\\
1+r_i-\frac{n+1}{2}
\;\;\;\;\;\;\;\;\;\;\;\;\;\;\;\;\;\;\;\;\;\;
\mbox{for} \;\; n\, \not\equiv \, 0 \, (2)
\end{array} \right.
\eeqa

{\em \underline{Remark:}} To give an example for an estimate of the $r_i$
we choose the base $B$ to be a del Pezzo surface ${\bf dP_k}$
(cf.~appendix \ref{app instanton}): here one has from the 
standing assumption\footnote{to have $C$ irreducible (and then $V$ stable)
one assumes in any case {\em just} that $\eta- nc_1$ is {\em effective}} 
(also adopted here by us), that $C$ is not only an effective divisor but 
even ample, that $\eta - nc_1$ is {\em even ample} 
and so $\eta \cdot E_i> n$ for the exceptional blow-up curves $E_i$. 
But there is, of course, a total of $240$ instanton curves on ${\bf dP_8}$, say,
not just the $8$ $E_i$.

Actually, however, one finds the derived bound for $r_i$ 
for {\em all} instanton curves in $B={\bf dP_k}$
and even more generally for all instanton curves in all bases $B$:
this follows from $K_B \cdot \C_i = -2 - \C_i^2$ and the fact that
all instanton (i.e. smooth isolated rational) curves $\C_i$ 
have selfintersection $\C_i^2=-1$ so that one has from $\eta - nc_1$ ample
that even in general
\beqa
\label{r_i bound}
r_i&> &n
\eeqa

The {\em conceptual identification} given in (\ref{vanishing locus}) of
a vanishing locus for the respective Pfaffians $Pfaff_{\C_i}$ for 
all the individual base instanton curves $\C_i$ now makes it possible
to {\em 'globalize'} (over the base $B$) the result immediately:
all individual Pfaffians can be made vanish {\em simultaneously} 
if one chooses the 
moduli\footnote{the coefficients of the $a_j$ in a polynomial language 
for $B$ a Hirzebruch surface ${\bf F_m}$, for example}
(globally over $B$ and not just over an individual $\C_i$)
to lie in the set $\R=\Si$ 
(where $\R\subset \cap_i \R_i $ obviously\footnote{\label{pullback footnote}
here obvious pullbacks
$res_i^{-1}\R_i$ are understood (with the restrictions
$res_i: \M_X(C)\ra \M_{\E_i}(c_i)$)})\\
{\bf \underline{Theorem}}  \fbox{$Pfaff_{all\; \C \subset B}$}\\
a) For an $SU(n)$ spectral bundle $V$ on an elliptic Calabi-Yau threefold $X$
one finds common zeroes for the Pfaffians
of all {\em base} curves $\C_i$ by choosing moduli from the set $\Si$
\beqa
\label{solution set}
\Si\; \subset \; \bigcap_{\C_i\subset B}(Pfaff_{\C_i})
\eeqa
b) one gets a suitable multiplicity $k\geq 2$ from (\ref{multiplicity}) 
applied to $r:=\mbox{min}_i \, r_i$ 
when one tunes the $r_i$ from a suitable choice of $\eta$: this needs
$r\geq 3$ for
the $SU(4)$ case and $r\geq 3$ or $4$ for the $SU(3)$ or $SU(5)$ case, 
respectively; from (\ref{r_i bound}) this is fulfilled automatically.

{\em \underline{Remarks:}}
a) Along the solution locus $\R$ the degrees of freedom in $\M_V$
given by the sections\footnote{they have class
$\eta-jc_1$; here we use the line bundle $\cL:=K_B^{-1}$ of class $c_1$} 
$a_j$ of $\N\ox \cL^{-j}$ over $B$ 
are, for $j=2, \dots, n-1$,
reduced to the freedom provided by sections $a_j^{\prime}$ of $\cL^{n-j}$
(they are of class $(n-j)c_1$)
in $a_j=a_j^{\prime}\cdot a_n$.
We compare the dimensions\footnote{assuming here that besides $\eta-nc_1$ 
also $c_1$ is ample, i.e.~$B={\bf dP_k}$, to use index computation} 
of the full moduli space $\M_V$ 
(reproducing (\ref{full moduli space dimension}))
and of our sublocus $\R$ 
(producing (\ref{full moduli space dimension}) 
with the coefficient $\frac{n}{2}$ of $\eta(\eta-nc_1)$ replaced by $1$): 
its codimension $(n-2)\frac{1}{2}\eta(\eta-nc_1)$
shows how much 
the partial moduli fixing has reduced the moduli freedom
(cf.~the remark 
concerning the situation for {\em one} instanton curve in app.~\ref{technical})
\beqa
\label{dim M_V}
dim \, \M_V&=& h^0\Big(B, \N\Big)+
\sum_{j=2}^{n-1} h^0\Big(B, \N\ox \cL^{-j}\Big) 
+ h^0\Big(B, \N\ox \cL^{-n}\Big)-1\;\;\;\;\;\;\\
dim \,\; \R&=& h^0\Big(B, \N\Big)\; +\;
\sum_{j=2}^{n-1}h^0\Big(B, \cL^{n-j}\Big)\; 
+ \; h^0\Big(B, \N\ox \cL^{-n}\Big)-1
\;\;\;\;\;\;\;\;\;\;\;\;\;\;\;\;
\eeqa
\indent
b) For $t\in \R$ the $z\in \M_{cx}$ 
remain flat directions (as do the $k\in \M_K$, 
cf.~remark in sect.~\ref{Pfaff special}): 
the conditions for $\R$
are posed in the $\M_V$ factor of $\M=\M_V\x\M_{cx}$ alone.

\subsection{\label{topological vanishing}Topological vanishing of the Pfaffian}

Sometimes one gets a vanishing result for a Pfaffian which is 
completely independent of the continuous moduli ('topological vanishing'). 
In that case the Pfaffian
vanishes along a whole component of the moduli space. These components
are labeled by the discrete degrees of freedom.
One such result is of course the identical vanishing of the Pfaffian (for
a base curve) for $SU(2)$ bundles 
mentioned in Theorem $Pfaff_{single \, \C \subset B}$ above.

We have focused above exclusively on the continuous moduli, so let us treat
here now also the discrete parameter $\la$ in (\ref{gamma class})
(and, to a certain extent, the discrete degrees of freedom 
encoded in the integral cohomology class $\eta$).
First and foremost one realises that the 
Theorem $Pfaff_{single \, \C\subset B}$
is completely independent of $\la$. The reason, of course, is that the 
degree of freedom incorporated in $\la$ just multiplies the object
whose triviality, along a certain locus in moduli space, is concerned
(the possible half-integrality of $\la$ turns out to play no role 
effectively).
So, here one gets a vanishing result for the Pfaffian which depends on
specific continuous moduli but is independent of the discrete modulus $\la$
(the conditions on $r_i=\eta\cdot \C_i$ were given above).

But there are also 'opposite' cases (so to speak)
where one gets a vanishing result for the Pfaffian 
which depends on specific choices of the discrete modulus $\la$ 
but is independent of the continuous moduli.
This happens if the effectivity of the divisor
$(\al s_i +\beta_i F)|_{c_i}$, related to the 
line bundle\footnote{for which one wants to establish the existence
of a nontrivial section (cf.~the step $\Si_i\subset (Pfaff_{\C_i})$ in the
proof of theorem $Pfaff_{single \, \C\subset B}$ in app.~\ref{proof appendix})}
$L|_{c_i}\ox \cO_{c_i}(-F|_{c_i})=\cO(\al s_i+\beta_i F)|_{c_i}$,
follows already from a corresponding effectivity of the divisor on $\E_i$,
not yet restricted to $c_i$. Clearly in such a case one gets a vanishing
result for $Pfaff_{\C_i}$ which is independent of the continuous moduli
and depends only on the discrete choices.

To find these cases
note first that one gets (from the first lines in (\ref{specialisation}))
from $\al\geq 0$ that $\la \geq - 1/2$, 
more precisely\footnote{as $\la$ is integral (and $r$ odd) 
for $n$ even and $\la$ is strictly 
half-integral for $n$ odd} that $\la\geq 0$ for $n$ even and $\la \geq -1/2$
for $n$ odd. One then implements in addition
the condition $\beta_i\geq 0$. This gives, as cases independent of $r_i$,
for $n$ even the case $\la=0$ and for $n$ odd the cases $\la= -1/2$ and $+1/2$
(note that these case were, for technical reasons, outside
the considerations of [\ref{BDO}] and [\ref{C1},\ref{C2}]).

Further inspection shows that there are in addition some exceptional cases.
Let us first recall that we have the restriction $r_i>n$
(from our assumption $C$ ample, cf.~app.~\ref{app instanton}). 
The condition $\beta_i\geq 0$ gives the
other bound $r_i \leq \frac{2\la n-1}{2\la -1}$ (we assume that we are now
in the $\la \geq 1$ regime). The universal case of $SU(2)$ was mentioned 
above.
The other cases of interest $n=3,4,5$, related
to an $E_6, SO(10), SU(5)$ GUT group (in the visible sector),
give then a list of exceptional cases

\noindent
{\bf \underline{Theorem 2}}  \fbox{$Pfaff_{all\; \C \subset B}$}\\
For an $SU(n)$ spectral bundle on an elliptic Calabi-Yau threefold 
the Pfaffians $Pfaff_{\C_i}$ of {\em all} base instanton curves 
$\C_i\subset B$ vanish {\em identically} in the following cases:\\ 
a)$\un{n \, \mbox{odd}:} \;\;\; \la =\pm 1/2$\\
\indent
$\un{n \, \mbox{even}:}\;\; \la =0$\\
b) Furthermore there are the following exceptional cases
(the $r_i$-conditions are for all $i$):\\
$\un{SU(2):} \;\; \mbox{universally}$\\
$\un{SU(3):} \;\; \la = 3/2$ and $r_i=4$\\
$\un{SU(4):}\;\; \la = 1$ and $r_i=5 \; \mbox{or}\; 7$\\
\indent\indent\indent\indent
$\la =2$ and $r_i=5$\\
$\un{SU(5):}\; \; \la = 3/2$ and $r_i=6 \; \mbox{or}\; 7$\\
\indent\indent\indent\indent
$\la = 5/2$ and $r_i=6$

\noindent
\un{\em Remarks:}
a) Note that by (\ref{la bound}) these cases are not so special
in $\la$ but just in $r_i$.\\
b) Often (if $B$ has many instanton curves) it will be difficult
to arrange that all the $r_i=\eta\cdot \C_i$ fulfil these conditions:
for example on the del Pezzo surfaces ${\bf dP_k}$ with $k\geq 5$ one has
(cf.~app.~\ref{app instanton}) as instanton curves the exceptional
blow-up curves $E_i$, $i=1, \dots , k$ and also curves of classes
$l-E_i-E_j$ and $2l-\sum_{n=1}^5 E_{j_n}$ which illustrates this problem. 

\subsection{\label{base relevance}
The relevance of vanishing results for base instanton curves}

In the last two subsections we found vanishing results for all individual 
{\em base} instanton curves, 
either for a certain subset among the continuous moduli 
(what, for abbreviation, we will call henceforth 'the special case'), 
or just specified by discrete parameters 
(what we will call 'the component case'). 
This does not give, of course, a satisfying answer
to our original question Pfaff$_{\scriptsize{\mbox{total}}}$(special).
This will be the reason for a second start in the next section
where we try to define bundles for which the corresponding argument can be 
made conclusive (i.e.~to include {\em all} instanton curves, 
not just those in a base).

Nevertheless let us point out the impact of the given partial results.
For this we have to isolate (as far as possible) in the general expression 
for the superpotential 
the directions belonging to the base and to the fibre, respectively.
So, if $[\C]=\sum n_jC_j+n_F F$ is the decomposition of the homology class of
an instanton curve in a homology basis (the $n_j$ and $n_F$ integral) and 
$J=\sum k_m J_m +k_F J_F$ a decomposition  
of a given K\"ahler class with respect to a basis 
dual\footnote{so $\int_{\C_j}J_m=\delta_{jm}, 
\int_F J_m=0, \int_{C_j}J_F=0, \int_F J_F=1$} 
to the $C_j$ (the $k_m$ and $k_F$ are the K\"ahler moduli), then one gets 
\beqa
W&=&\sum_{l=0}^{\infty}\, a_l\, q^l\; = \; W_B+W_{skew}
\eeqa
(with $q:=e^{ik_F}$ and $W_B=a_0$)
where $a_l=\sum_C P_C e^{i\sum n_j k_j}$ and the summation over $C$ goes over 
classes which have fibre component part $n_F F=lF$. 

Now, what the results up to now say,
is just that the constant term (in $q$) of this series 
can be made to vanish: $a_0=0$.
More precisely, this happens  
either for $t\in \R$ (with $\la$ arbitrary; this was our 'special case') 
or for one of the whole 
components (codimension zero) $\M_X(C)$ of the bundle moduli space 
specified by the discrete parameters as in 
Theorem 2 $Pfaff_{all \, \C \subset B}$ above (the 'component case').
What one learns from this is just that, at a certain 
boundary\footnote{so this will be in any case 
only a 'solution' in a generalised sense}
region of $\M_K$, 
more precisely for the partial decompactification limit of large fibre,
one has $W=0$ (this holds {\em with the 
bundle moduli specified} as in the special case or the component case);
note that one has for this region {\em in any case} $\partial_{k_F}W=0$.  
Moreover one has in this region, in both cases (special or component), 
that $\partial_{k_j}W=0$
(again the only nontrivial part of this assertion is the behaviour of the 
constant term where the assertion follows directly from our theorems above).

Thus the mentioned theorems (in both cases) provide indeed supersymmetric 
Minkowski vacua, but only in a 'generalised sense' 
because one is, in this partial decompactification limit,
not in the bulk but in the boundary of the K\"ahler moduli space. 

\subsection{\label{location subsection}
The location of the solution set in the moduli space}

We now will have a closer look on the actual solution set, the
subset $\R= \Si$ of $(Pfaff)$, and its location in the moduli space $\M_V$.
Recall that the latter has 
$\M_X(C)\cong {\bf P}H^0(X, \cO_X(C))\cong |C|$ (the linear system)
as its continuous part and the discrete choices involved in $\eta$ and $\la$
as labels for the components.

Let us recall first that indeed the discrete choice for the 
half-integral number $\la$ labels distinct branches of the moduli space.
This is in a sense obvious but becomes also particularly palpable from
the following perspective:
the other discrete choice, that of the class $\eta$, will lead in general
to $\eta(\eta-nc_1)\neq 0$; in such a case it is clear from 
$\frac{1}{2}c_3(V)=\la \eta(\eta-nc_1)$, cf.~[\ref{C}], that indeed
bundles constructed from different $\la$ lie in different branches 
(components disconnected to each other) of the moduli space $\M_V$.

Now let us look what happens to the line bundle $L$ on $C$, from which
$V$ is constructed as $p_*(p_C^* L \ox \cP)$, along the specific locus $\Si$ 
in the moduli space $\M_X(C)\cong {\bf P}H^0(X, \cO_X(C))$ where
$\cO_X(\widetilde{\Lambda})|_C=\cO_X\Big(n\si-(\eta-nc_1)\Big)\Big|_C$ 
becomes trivial.
There one has from 
\beqa
L&=&
\cO_X\Bigg(n(\la + \frac{1}{2})\si 
+\Big[(n\la+\frac{1}{2})c_1-(\la - \frac{1}{2})\eta\Big]\Bigg)\Bigg|_C
\eeqa
the following specialisations\footnote{Recall that for $n$ even one has
$\la$ integral and $\eta\equiv c_1\, (2)$ 
as the case of a strictly half-integral
$\la$ is excluded (it would need an even $c_1$ which itself is excluded by
$\chi_i=\C_i\cdot c_1=1$).}
\beqa
\label{Sigma specialisations}
L&\stackrel{\Sigma}{\ra}&\left\{ \begin{array}{ll}
\cO_{X}\Big(\frac{n}{2}\si+\frac{\eta +c_1}{2}\Big)\Big|_C
& \;\;\;\;\;\;\;\mbox{for} \; n \, \mbox{even} \\
\cO_{X}\Big(\eta -\frac{n-1}{2}c_1\Big)\Big|_C
& \;\;\;\;\;\;\;\mbox{for} \; n \, \mbox{odd} 
\end{array} \right.
\eeqa

Therefore, regardless of their specific form, {\em the dependence on the 
parameter $\la$ has dropped out completely} !
This would mean that (for fixed $n$ and $\eta$) all the 'distinct'
components with continuous part $\M_X(C)$ 
and just different discrete labels $\la$
are connected via the subset $\Si$ in which they meet. This, clearly, can
mean only that $\Si$ actually does not lie in the bulk of $\M_X(C)$
but rather in its boundary. Therefore the supersymmetric solutions
to the system $DW_B=W_B=0$ (where $W=W_B+W_{skew}$)
we did find are solutions only in a generalised sense
as they lie on a 'degeneration boundary' 
of $\M_X(C)$ (itself the continuous part of $M_V$), 
which, however, is not an uncommon thing.

Let us try to understand this phenomenon more directly: from a
degeneration of $C$, enforced when going to $\Si$. For this we
concentrate on the explicit description $\R$ of $\Si$ 
which contains directly the
defining coefficient sections $a_j$ of the spectral cover equation and 
is therefore most accessible for 
an investigation which wants to find out the singularity of the 
corresponding spectral surfaces.

It is enough to illustrate what happens for the three cases
of greatest interest, i.e.~the $SU(3), SU(4)$ and $SU(5)$ bundles.
They have respective {\em affine} cover equations 
$w_5=a_0+a_2 x+a_3 y + a_4 x^2 + a_5 xy$, with $w_4$ (or $w_3$) 
arising by setting $a_5$ (or $a_5 $ and $a_4$) to zero. So one has 
(the mentioned specialisations give the corresponding results 
for $w_3$ and $w_4$)
\beqa
\p_x w_5&=&a_2+2a_4+a_5y\\
\p_y w_5&=&a_3+a_5 x
\eeqa
The surface $C$ is the intersection\footnote{inside 
the fourfold given by the 
${\bf P^2}$ bundle, with projective coordinates $(x,y,z)$, over $B$}
of the elliptic Weierstrass 
equation and the spectral cover equation $w=0$. 
So along the locus $\R$ in moduli space
the expression $w$
(in the affine patch described) has vanishing gradient at $\pi_C^{-1}(A)$
where $A=C\cap \si$ is the vanishing divisor of $a_n$ in $B$ 
(cf.~\ref{zeroes in the base}; $\pi_C=\pi|_C$), 
so $C$ is singular there. As explained in the 
remark in app.~\ref{Characterisation} 
one has here $\pi_C^{-1}(A)=nA$, so $A$ is a 
total ramification locus for $\pi_C$.

\noindent
\un{\em remark:} Two all-important caveats: 
1) a comprehensive degeneration dictionary between 
$C$ and $V$ has not been worked out; 2) we {\em assume} that our
reasonings remain valid despite the degeneration
(often this is so by continuation from the interior of the moduli space).

So, remarkably, (for fixed $n$ and $\eta$) 
this singular degeneration of $C$,
which lies at a certain boundary location of the continuous part
$\M_X(C)$ of the multi-component space $\M_V$, 
{\em connects all the different components labeled by the parameter $\la$}
(as being common to all of them, cf.~after (\ref{Sigma specialisations}); 
one convinces oneself 
also about the opposite inclusion)\footnote{\label{caveat footnote}Many 
statements throughout the paper, like whether $t\in \M_X(C)\subset \M_V$
or actually $t\in \ov{\M_X(C)}\subset \ov{\M_V}$, 
have to be read with this bulk versus boundary question in mind; only in the 
present subsection did we make this distinction notationally manifest. 
The question of when (i.e.~for which modulus) 
and where (in $C$) a spectral object $C$ becomes singular
has to be investigated case by case: the distinction 
(like $\Si\subset \ov{\M_X(C)}$, with $C$ singular at $A$ for $t\in \Si$)
could then be made manifest also in the corresponding statement.
Also all statements have to be read with the assumption of the
caveat remark above in mind.}
\beqa
\Si_{n, \eta}&=&\bigcap_{\la} \, \ov{\M(V)_{n, \eta}^{\la}}
\eeqa
\indent
In particular, at this locus 
in the {\em common} (!) boundary of all
the $\la$-components the amount of chiral matter can be
arbitrarily switched (in units of $\eta ( \eta - n c_1)$). 
This {\em chirality change phenomenon} should not come as a surprise because
the chiral matter is known [\ref{C}] to be related to the curve $A$.
That is, when going to this common boundary locus, all the different
generation numbers labeled by $\la$ jump so that - {\em via going through
this singular configuration - all generation numbers are connected}
(which might have interesting consequences for phenomenology;
at the same time this 'jumping locus' is singled out as a specific 
supersymmetric solution, though just for the $W_B$ part and only 
in the described generalised sense as it lies at a degeneration boundary).

\newpage

\section{\label{new class}A new class of bundles}

\resetcounter

As we have seen, our approach - to search for critical points of the 
world-sheet instanton superpotential $W$ 
by searching for critical points of the individual Pfaffians for all
contributing instanton curves - 
did have, when applied to spectral cover bundles 
on elliptically fibered $X$, only partial success: 
only the {\em base} instanton curves $\C_i\subset B$
were under control. Therefore we now make a fresh start and define
a new class of bundles, now on an {\em arbitrary} Calabi-Yau threefold
(not necessarily elliptically fibered), where we want
to extend the previous method
to a set-up where ${\em all}$ instanton curves are under control.

\subsection{The idea of the construction}

The idea of the construction to follow is, of course, to use the control
which one has over the Pfaffians for base instanton curves, 
in the spectral cover approach for spectral bundles over 
elliptic Calabi-Yau spaces $X$, now for all instanton curves in an
arbitrary, not necessarily elliptically fibered $X$. For this let us first note
that in sect.~\ref{first try} 
{\em the whole elliptic fibre direction played only a rather marginal role}: 
we considered only base instanton curves $\C_i\subset B$
and so were only interested in the behaviour of the bundle when restricted
to $B$ where the relation (or its versions 
restricted to the situation over a curve $\C_i$) 
\beqa
\label{base relation}
V|_B\cong \pi_{C*}L 
\eeqa
was crucial.
The whole fibre direction, 
i.e. the whole manifold $X$ beyond just the base $B$,
was only important as an ambient space where a cover manifold $C$ of $B$
lives. The line bundle $L$ on $C$ was essentially fixed by our condition 
$c_1(V)=0$. The ambient space $X$ was then important for an explicit 
description, via the equation (\ref{C equation}), 
of the possible concrete incarnations of $C$ in its linear system.
This allowed to describe the (continuous) moduli of $V$ directly
as $\M_X(C)\cong {\bf P}H^0(X, \cO_X(C))$.

Therefore it should be clear which parts of the spectral cover construction
are actually responsible for its (partial) success: when controlling the
Pfaffians for base instanton curves $\C_i\subset B$ one does not need
to have any information about the behaviour of the bundle $V$ on $X$ 
outside of $B$.
The fibrewise description of $V$ starting from the Poincare line bundle 
is not needed. The only impact of $C$ is described in 
(\ref{base relation}).

So, if we want to have control over {\em all} instanton curves in 
an arbitrary (not elliptic) $X$, we just need to consider
the bundle $V$ on $X$ as such a base part of a set-up in a fictitious
elliptic fourfold $Y$ (not Calabi-Yau) over $X$ where we just demand
that our bundle arises as a projection (a push forward) from a line 
bundle on a (threefold) cover $C$ as in (\ref{base relation}).
The ambient space $Y$ serves only as an auxiliary space to have a class
of possible covers $C$ under control; i.e., the motions of $C$ in $Y$
will map to the moduli of $V$ on $X$.

\subsection{\label{new classes}New classes of bundles}

Let us first explain our emphasis in the 
notion of what is 'new' with defining the classes of bundles below.
The possibility to define such bundles is not new. 
The point is that we single out these bundles as objects to be studied
because it turns out to be possible to apply the ideas of the previous section
to gain control now over {\em all} Pfaffians;
this gives these classes certainly a special importance.

One may give the relevant definitions first quite generally 
in the category of smooth compact complex Kahler manifolds 
(one may also think already of the algebraic category)
of arbitrary dimension $k$ and holomorphic bundles.

\noindent
\underline{\em {\bf Definition 'Projected Bundles'}} \\
A $U(n)$ vector bundle $V$ on an $k$-fold $X$ is called 'projected' 
if there is a (ramified) covering
$k$-fold $p_C: C_k\ra X$ and a line bundle $L$ on it 
such that $V\cong p_{C*} L$.

One would also like to know whether a {\em given} bundle is of this type; 
this leads to the question of giving an internal characterisation 
of a bundle $V$ over $X$ to be projected.

If $n$ is the covering degree of $p_C:C\ra X$, such that $rk(V)=n$, 
one finds\footnote{with the
Grothendieck-Riemann-Roch formula $ch(V) Td(X)=p_*(ch(L) Td(C))$}
for a projected bundle $SU(n)$ bundle
(i.e.~for simplicity we assume 
$c_1(V)=p_*(l+\frac{1}{2}c_1(C))\, \buildrel !\over = \, 0$)
\beqa
\label{equ c_2(V)}
c_2(V)&=&-p_*\Bigg( \frac{1}{2}l^2+\frac{1}{2}l \, c_1(C)
+\frac{c_2(C)+c_1^2(C)}{12}\Bigg) \\
\label{equ c_3(V)}
c_3(V)&=&p_*\Bigg( \frac{1}{3}l^3+\frac{1}{2}l^2 \, c_1(C)+
l\, \frac{c_2(C)+c_1^2(C)}{6}+\frac{c_1(C)c_2(C)}{12}\Bigg)\;\;\;\;
\eeqa
(where $l:=c_1(L)$ and $p:=p_C$).
Usually one will use these results in a slightly different form where one makes
manifest the condition $c_1(V)=0$ by the following choice of $l$
\beqa
\label{l specification}
l&=&-\frac{1}{2}c_1(C)+\ga_3
\eeqa
Here $\ga_3$ is assumed to be in the 
kernel of $p_{*}:\, H^{1,1}(C)\ra H^{1,1}(X)$.

In the situation where the topological type of $C$ is fixed
(for more discussion about the possible variation of the cover threefold
$C$ cf.~below) all the dependence of the Chern classes on the input 
parameters of the construction lies therefore in this class $\ga_3$.
One finds then (where we use the abbreviations $c_i:=c_i(C), \ga:=\ga_3$)
\beqa
c_2(V)&=&-\frac{1}{2}p_*\Bigg( \ga^2 + \frac{2c_2-c_1^2}{12}\Bigg)\\
c_3(V)&=&p_*\Bigg(\frac{1}{3}\ga^3+\frac{2c_2-c_1^2}{12} \ga\Bigg) 
\eeqa
Further specification of the classes $c_i$ and $\ga$ will arise in the
constructions defined below.

{\em Refined constructions}

To have better control over the possible $C_k$, that is to describe their
possible moduli more explicitly, 
we will also consider some refinements. For this note first that one may
consider our definition in loose analogy to another class of bundles,
introduced also in connection with the question to gain control over the 
contribution from the Pfaffians of all the world-sheet instantons [\ref{BW}].
There one has also a second space $Z$, endowed with a bundle $V_Z$, 
and a map $i: X\ra Z$. From these data the given bundle $V$ on $X$ is derived
\beqa
i: X&\ra &Z\\
V&\cong & i^*V_Z
\eeqa
Here $i$ is an embedding of $X$ as a hypersurface (or complete intersection)
in $Z$ and $rk\, V = rk\, V_Z$.
In our (in a sense 'dual') case one has
\beqa
p: C&\ra & X\\
V&\cong & p_* L
\eeqa
Here $p$ is a projection which is a finite (ramified) covering
and $rk\, L=1$.

To make actual progress with the investigation of a projected bundle $V$ 
one has to assume more structure on the auxiliary space, i.e.~$Z$ in [\ref{BW}]
and $C$ here: just as $Z$ is assumed to be a toric variety, here $C$ will
be assumed to be describable as follows

\noindent
\underline{\em {\bf Definition 'Embedded Projected Bundles'}} \\
A projected bundle $V$ on $X$ is called 'embedded projected' 
if there exists an embedding $i_C: C_k\ra Y_{k+1}$ of $C_k$ 
in an $k+1$-fold $Y$

(Again one may try to characterise this class of bundles more intrinsically.)
Now, often one will be in a situation where the line bundle $L$ on $C$ is fixed
(up to discrete degrees of freedom)
already by its cohomology class (the first Chern class) and has no further 
continuous moduli. In that case the moduli come from the possible motions 
of $C$ in $Y$ (i.e.~the possible actual hypersurfaces in the given linear 
system). 
So one gets then a projection 
$\M_Y(C)\cong {\bf P}H^0(Y, \cO_Y(C))\ra \M_X(V)$.

To get further information on the degrees of freedom of the hypersurface $C$
in $Y$ one puts additional structure on $Y$, like the following
(needless to say, again it would be of interest to characterise the class 
intrinsically)

\noindent
\underline{\em {\bf Definition 'Fibered Embedded Projected Bundles'}} \\
A projected bundle $V$ on $X$ is called 'fibered embedded projected' 
if there exists an embedding $i_C: C_k\ra Y_{k+1}$ of $C_k$ 
in an $k+1$-fold $Y$ which has a fibration $\pi: Y\ra X$

We will usually add here the condition that the fibration has a section
and $X$ is considered as divisor in $Y$.
A we consider $C$ in the embedded case just as a 
hypersurface in $Y$\footnote{in the following 
we suppress the dimension subscript 
and write just $C$ and similarly $p$ for $p_C=\pi|_C$}
one has then, in the generic case (which we assume) where just $X$ and $\pi^* S$
are available,
for the cohomology class\footnote{if no confusion 
can arise we denote the cohomology class
of $C$ or $X$ in $Y$ just by $C$ and $X$, resp.; 
corresponding notations will be used below} 
of the divisor $C$ in $Y$ (with $S\in H_4(X)$)
\beqa
C&=&nX+\pi^*S
\eeqa

\subsection{Elliptically embedded projected bundles}

In the class of fibered embedded projected bundles one may distinguish
according to the genus of the fibre: a ruled case, an elliptic case
and so on. 
We take as the class which is related most directly to the
set-up we have in mind the following

\noindent
\underline{\em {\bf Definition 'Elliptically Embedded Projected Bundles'}} \\
A projected bundle $V$ on $X$ is called 'elliptically embedded projected' 
if there exists an embedding $i_C: C_k\ra Y_{k+1}$ of $C_k$ 
in an $k+1$-fold $Y$
which is elliptically fibered $\pi: Y \ra X$ over $X$ with $X$ as a section.

Again one may ask for an internal understanding of this class, that is,
in refinement of the earlier characterisation issue,
one would now 
like to have an internal characterisation of a bundle $V$ over $X$ 
to be elliptically embedded projected.

Concretely $Y$ will be described by a 
Weierstrass equation\footnote{with $x,y,z,g_2,g_3$ sections of $\cL^k$ for
$k=2,3,0,4,6$; the  $x,y,z$ being homogeneous coordinates of ${\bf P^2}$}
$zy^2=4x^3-g_2xz^2-g_3z^3$ in a ${\bf P^2}$
bundle ${\bf P}(\cL^2\oplus\cL^3\oplus \cO)$ over $X$ where $\cL$ is a line
bundle over $X$ (which we assume to be nontrivial;
{\em note that no bundle on $Y$ does occur here, 
no fibre product, no Poincare bundle}).
The hypersurface $C\subset Y$ will be described by an equation
(for $n$ even, say) 
\beqa
\label{the C equation}
w=a_0+a_2 x + a_3 y + \dots + a_n x^{n/2}=0
\eeqa
(using affine $x,y$)
with the $a_i$ sections of $\cO_X(S)\ox \cL^{-i}$. 
We will use the abbreviations
\beqa
c_1(L)&=&l\in H^2(C)\\
\label{la definition}
c_1(\cL)&=&\lambda\in H^2(X)
\eeqa

Clearly, the restrictions to the base $B_2$ or to a 
curve $\C$ in it of a spectral cover bundle in an elliptic Calabi-Yau 
threefold are examples of embedded projected bundles. What is new is that
we will apply now this set-up to the bundle $V$ on our arbitrary Calabi-Yau
threefold $X$ (which is our case from now on) as a whole.

For an elliptically embedded projected bundle
one finds\footnote{note that $X^2=-\la X$; 
here the right hand sides in the expressions for $c_k(C)$
are - with some suppressed pullbacks from 
$X$ to $Y$ included - in $H^*(Y)$, and then restricted to $C$} 
from the equation for $C$
\beqa
\label{equ c_1(C)}
c_1(C)&=&-\Big(nX+S+\la\Big)\Big|_C \\
\label{equ c_2(C)}
c_2(C)&=&\Big(c_2(X)+12\la X+11\la^2+(nX+S)^2+\la(nX+S+\la)\Big)\Big|_C 
\eeqa

Taken together (\ref{equ c_2(V)}), (\ref{equ c_2(V)}) 
and (\ref{equ c_1(C)}), (\ref{equ c_2(C)}) give 
necessary conditions on $c_k(V)$ for an $SU(n)$ bundle 
$V$ to be elliptically embedded projected (as we assume from now on).

The degree of freedom introduced by the twisting line bundle $L$ on $C$
can be described more explicitly: one finds
(cf.~(\ref{l specification}); 
suitable pullbacks to $Y$ for $S$ and $\la$ understood)
\beqa
\label{c_1(L) representation}
l&=&\frac{nX+S+\la}{2}\Big|_C +\ga_3
\eeqa
where $\ga_3$ is in the kernel of $p_{C*}: H^{1,1}(C)\ra H^{1,1}(X)$.
Generically - {\em and we will assume that this generic class 
is the one actually used} - this $\ga_3$ will be just a multiple, $\mu$ say,
of $nX-(S-n\la)$ (understood as class in $Y$ restricted to $C$); 
here $\mu\in \frac{1}{2}{\bf Z}$ with actually $\mu \in \frac{1}{2}+{\bf Z}$
for $n$ odd, while for $n$ even $\mu \in \frac{1}{2}+{\bf Z}$ needs
$\la$ even and $\mu \in {\bf Z}$ needs $S+\la$ even. From here one can write
$L$ (even uniquely for $H^{0,1}(C)=0$, which itself would follow from 
$H^{0,1}(Y)=0$ and $C$ ample in $Y$) as a line bundle $\cO_Y(D)|_C$.

The genericity assumption made here concerning the concrete form of the 
twist class $\ga_3$ will turn out to be so important that we want to state it
explicitly

\noindent
\underline{\em {\bf Definition 
'Generic Elliptically Embedded Projected Bundles'}} \\
An eliptically embedded projected bundle $V$ is called generic
if the twist class $\ga_3$ of the line bundle $L$ on $C$ is built
from the generic divisors only.

These generic divisors on $Y$ are $X$ (embedded as section) and the pullbacks
$\pi^*S$ for $S$ a divisor in $X$.
As the concrete form of the twist played a crucial role in the proof of 
Theorem $Pfaff_{single \, \C \subset B}$, cf.~app.~\ref{proof appendix},
on which our generalization will be modelled, we restrict us in the following 
to this generic class (in the described sense; 
cf.~also sect.~\ref{relevance of genericity}).

\newpage

It will be important 
to determine the region in the Kahler cone where an embedded projective
bundle is stable, in particular whether this region is 
nonempty\footnote{
This will be sufficient in our set-up: we do not have to care that the 
stability region of $V$ in $\M_K$ includes specific Kahler moduli values
which are potentially forced on us by moduli fixing from $W$; for in our
procedure the vanishing of all the $P_C$ (even of all individual Pfaffians),
at the locus in $\M$ we are going to describe, clearly leaves the Kahler
moduli completely unfixed as remarked earlier on.}.

\subsubsection{An example}

In the special case that $X$ itself is actually elliptically fibered 
(with section) there is an important special class of bundles, 
the spectral bundles, which can be characterised abstractly
as being fibrewise semistable (concretely: 
a sum of line bundles of degree zero).
Such bundles have a non-empty stability region in the K\"ahler
moduli space $\M_K$. They turn out to be
of elliptically embedded projected type, but are not generic in our sense here
as their twist line bundle involves the Poincare bundle which itself
involves the diagonal class which is a non-generic divisor
(in the sense used here; cf.~sect.~\ref{relevance of genericity}).  

\noindent
\underline{\em {\bf Fact}} \\
On an elliptically fibered Calabi-Yau threefold $X$ (with section)
the spectral bundles are examples of elliptically embedded projected bundles 
$V$. They are stable but not generic.

So consider an elliptic $X$, with fibration $\pi_B:X\ra B$ of fibre $F$ 
and section $\sigma$ and with a spectral bundle $V$
(with spectral surface $C_2\subset X$ of class $n\sigma+\pi_B^* \eta$, 
endowed with a line bundle $L_2$).
For such a bundle we use\footnote{actually $C_3$ has singular points which are 
discussed in [\ref{C}] together with a resolution one can use} 
$C_3:=X\x_B C_2$ (of class $nX_1+\pi^*(\pi_B^* \eta)$), $Y:=X\x_B X$
and the line bundle on $C_3$ (with the Poincare bundle $\cP$ on $Y$, cf.~below)
\beqa
\label{L_3 definition}
L_3&:=&p_{C_2}^*L_2\ox \cP
\eeqa 
Correctly interpreted,
the equation for $C_3\subset Y$ is the same as the one for $C_2\subset X$
and $\M_Y(C_3)\cong \M_X(C_2)$.
Choosing $C_2$ ample in $X$ makes $C_3$ ample in $Y$.
Considering the first factor $X_1$ in $Y$ as the Calabi-Yau space $X$
we actually consider, the elliptic fibration from the second factor
$X_2$ in $Y$ is the auxiliary fibration (over $X_1$)
of the ambient space $Y$ of $C_3$. The cohomology class $\la$
of (\ref{la definition}), related to this fibration $\pi=\pi_1$
of $Y$ over $X_1$, is the same as for the fibration of $X_1$ over $B$: 
it is the class $\pi_B^* c_1$.

\subsubsection{\label{relevance of genericity}
On the relevance of the genericity assumption}

The definition of genericity given earlier may seem natural enough;
nevertheless the proper importance of this concept in our investigation
lies in the following reasoning.
Our strategy to gain control over {\em all} instanton curves on a general $X$
is to adopt the proof of theorem
$Pfaff_{single \, \C\subset B}$ (which gave control over the contribution
of {\em base} instanton curves for a spectral bundle).
When one goes through the analogy on which this procedure relies 
(cf.~the only slight deviations in app.~\ref{deviations} compared to
app.~\ref{proof appendix}; just a curve-dependent $\chi_i$ 
is implemented in addition) one finds that
the argument for $\R_i\subset \Sigma_i$ is generally true
but the second step $\Sigma_i\subset (Pfaff_{\C_i})$ makes use of the 
concrete form of $L|_{c_i}$. This in turn came from the representation
(\ref{c_1(L) representation}), more precisely from the concrete form
of the gamma class $\gamma=\mu (nX-(S-n\la))|_C$
{\em whose triviality (when restricted to $c_i$) 
is assured along $\Sigma_i$ in moduli space} 
(meant is: the corresponding {\em divisor class} on $c_i$
becomes linearly equivalent to zero along $\Sigma_i$;
this leaves the line bundle of an effective divisor,
having a non-trivial section, thus concluding the argument).
The form of $\ga$ arose because the 
divisors used ($X=X_1$ and $\pi^{-1}S$, for $S\subset X$ a divisor)
are the only ones available in general.

However, the concrete form (\ref{L_3 definition}) above, related to 
the divisor $p_{C_2}^*c_1(L_2)+\Delta-X_1-X_2+K_B$ 
shows that in $p_{C_2}^*c_1(L_2)$ all classes are of the desired form
but in $c_1(\cP)=\Delta-X_1-X_2+K_B$ (as divisor) 
only the classes $X_1$, $X_2=\pi^{-1}B$ 
and $K_B$ (which means of course $\pi^{-1}\pi_B^{-1}K_B$) are of this form; 
the diagonal $\Delta$ is not of this type.
Comparing (\ref{c_1(L) representation}) and (\ref{L_3 definition})  
one has $\ga_3=p_{C_2}^*\ga_2+c_1(\cP)$.
For $p_{C_2}^*\ga_2$ the usual argument applies: it is 
made linear equivalent to zero when $C_3$ is tuned in its moduli space
appropriately (from tuning $C_2$ in {\em its} moduli space). 
The remaining terms, however, in particular the diagonal, show that our 
assumption about the form of $c_1(L)$, expressed after (\ref{c_1(L) 
representation}), is violated. 

Moreover it is not possible to extend the original argument to include
this case as one can not tune $C_3$
such that $c_1(\cP)|_{C_3}$ (as divisor) becomes linearly equivalent 
to zero: one has $c_1(\cP)\cdot C_3=c_1(\cP)\cdot \eta$, and intersecting  
with a suitable further class $\pi^{-1}\pi_B^{-1}\psi$
gives the class $\Delta-F_1-F_2$ (over all points
of $\eta \psi$ in the respective fibre surface\footnote{the product of 
the two elliptic fibers $F_1$ and $F_2$; 
we use the same symbol for various diagonal classes});
intersecting this further with $\Delta$ gives $0-1-1\neq 0$ 
(as $e(\Delta)=0$ for the elliptic diagonal).

\subsection{\label{moduli of proj}The moduli space and the solution set}

The mentioned twist-parameter $\mu$ gives a {\em discrete} degree of
freedom in the construction. 
Given the way how $V$ is defined as embedded projected bundle
its {\em continuous} moduli come from the deformations of $C$ in $Y$, 
i.e.~from the choice
of the concrete equation (\ref{the C equation}) for $C$. 
So, keeping $\mu$ fixed,
one has a map\footnote{had we constructed a spectral cover bundle 
$V_Y$ over $Y$ from $C_3$ with Poincare bundle over $Y\x_X Y$,
$P$ would be the restriction map $\M(V_Y)\ra \M(V)$ 
(concerning the continuous moduli) as then $V=V_Y|_X$.} 
\beqa
\label{moduli surjection}
P: \M_Y(C)\cong {\bf P}H^0(Y, \cO(C))&\ra & \M(V)
\eeqa
The moduli $v_i$ of $V$ are
an image of the moduli $c_j$ of $\M_Y(C)$ under $P$.

This means that when it comes to taking derivatives\footnote{we may assume 
that we pose also the condition $W=0$ bringing us back from
covariant derivatives (including the Kahler potential) to ordinary derivatives}
of the superpotential $W$ that one has from 
$\frac{\p W}{\p c_j}=\sum_i \frac{\p v_i}{\p c_j} \frac{\p W}{\p v_i}$
only an implication in the wrong direction: from having $\p_{v_i}W=0$
to $\p_{c_j}W=0$ instead of the opposite what would be the implication we want;
for our goal is the relation $\p_{v_i}W=0$ and where we have control is
$\p_{c_j}W=0$. However, the surjection (\ref{moduli surjection}) 
tells us in particular that there are
'more' $c_j$ moduli than $v_i$ moduli. With our control over the $\p_{c_j}W$
we might have enough information to solve our proper problem.
Assume a moduli choice $\{c^*_j\}$ in a solution set in $\M_Y(C)$ solves
$\p_{c_j}W=W=0$ and let $\{v^*_i\}$ be the image under (\ref{moduli surjection})
in the image of the solution set; clearly this will also solve $W=0$.
In the relation $(\p_{c_j}W)_{j}=A\cdot (\p_{v_i}W)_{i}$ 
the $j$-index runs over $1, \dots, p$ and the $i$-index over 
$1, \dots, q$ where $p\geq q$; so rank $A \leq q$. Choose local coordinates
such that only the first $q$ lines of $A$ are non-zero, 
constituting a matrix $\A$;
under the (generic) assumption rank $\A=q$ we can invert the relation
of derivatives and have then the possibility to make an implication in the
direction we actually want.

The rest of the argument for having 
vanishing Pfaffians (at certain moduli), now for {\em all}
instanton curves in $X$, runs in parallel to the earlier set-up.
If one specialises the $c_j$-moduli to the set (\ref{X solution set}) below
one finds that the Pfaffians for all instanton curves $\C$ in $X$ will vanish
at the corresponding $v_i$-moduli; again one can include higher multiplicities
from tuning $S$ suitably 
such that $r:=\mbox{min}_i \, r_i$ is suitably large where
\beqa
r_i&:=&S\cdot \C_i\\
\chi_i&:=&\la\cdot\C_i 
\eeqa 
Actually one now is in a set-up with different $\chi_i$
for each $i$ (cf.~app.~\ref{app instanton} and app.~\ref{deviations}).
Under suitable
assumptions, like maximal rank (of the differential) 
of the mapping $P$, as described above, 
and for $\la$ ample\footnote{and not being even, 
for simplicity, cf.~the remark after (\ref{specialisation in new set-up});
also, for simplicity, we did assume $h^{1,0}(Y)=0$ and $C$ ample in $Y$}, 
one finds 
(cf.~app. \ref{deviations} for the proof;
furthermore here the notation concerning the moduli space 
has to be read with footn.~\ref{caveat footnote} in mind)\\ 
{\bf \underline{Theorem}}  \fbox{$Pfaff_{all\; \C \subset X}$}\\
a) For a generic elliptically 
embedded projected $SU(n)$ bundle $V$ on an arbitrary Calabi-Yau threefold $X$
one finds common zeroes for the Pfaffians $Pfaff_{\C_i}$
of {\em all} instanton curves $\C_i$ by choosing the bundle moduli from 
the image under $P$ in (\ref{moduli surjection}) of the set 
\beqa
\label{X solution set}
\R \; := \; \Big\{ t\in \M_{Y}(C)\Big| \; \;
a_{n} | a_{j}\;\; \mbox{for}\;\; j=2, \dots , n-1\Big\}
\eeqa
b) one gets a suitable multiplicity $k\geq 2$ for all $\C_i$.

\subsection{Topological vanishing of the Paffian}

For completeness we should point out that here again there will be cases
where one gets vanishing results for Pfaffians which are completely independent 
of the continuous moduli, i.e.~where a Pfaffian vanishes along a whole
component of the moduli space ('topological vanishing').

To get such results one adopts the same strategy 
as in sect.~\ref{topological vanishing}: one searches for 
the numerical conditions
on the parameters $\mu\in \frac{1}{2}{\bf Z}$ and $r_i\in {\bf Z}$ (where 
now under our assumptions, cf.~app.~\ref{deviations}, one can assume that 
$r_i>n\chi_i$ and $\chi_i>0$) 
which assure the effectivity of the divisor of the relevant
line bundle on the respective $c_i$, again already from effectivity
of the corresponding divisor on $\E_i=\pi^{-1}c_i$
(cf.~for these divisors 
the first lines in (\ref{specialisation in new set-up})).

We give again the exceptional cases for the 
cases of interest, i.e.~$SU(3), SU(4), SU(5)$ bundles which lead to
corresponding GUT groups in the (visible) unbroken gauge group.

\noindent
{\bf \underline{Theorem 2}}  \fbox{$Pfaff_{all\; \C \subset X}$}\\
For a generic elliptically embedded projected $SU(n)$ bundle on an 
arbitrary Calabi-Yau threefold 
the Pfaffians $Pfaff_{\C_i}$ of {\em all} instanton curves 
$\C_i\subset X$ vanish {\em identically} in the following cases:\\ 
a) First the general cases (independent of the $r_i$):\\
\vspace{.1cm}
\noindent
$\un{n \; \mbox{odd}:} \;\;\; \mu =\pm 1/2$\\
\vspace{.2cm}
\noindent 
$\un{n \; \mbox{even}:}\;\; \mu =0$\\
\vspace{.2cm}
b) Furthermore there are the following exceptional cases 
(now $\mu > 1/2$):\\
\vspace{.2cm}
\noindent
$\un{SU(2):} \;\; \mbox{universal}$\\ 
\vspace{.3cm}
\noindent
$\un{SU(3):} \;\; 
3\chi_i\; <\; r_i\; \leq \;
3\chi_i+\frac{2\chi_i-1}{\mu-\frac{1}{2}}$\\
\vspace{.3cm}
\noindent
$\un{SU(4):}\;\; 4\chi_i \; <\; r_i\; \leq \;
4\chi_i+\frac{\frac{5}{2}\chi_i-1}{\mu-\frac{1}{2}}$\\
\vspace{.3cm}
\noindent
$\un{SU(5):}\; \; 5\chi_i\; <\; r_i\; \leq \;
5\chi_i+\frac{3\chi_i-1}{\mu-\frac{1}{2}}$

As in the previous case of such a theorem (for just the base curves of an
elliptic Calabi-Yau space) here again in the exceptional cases
it will be usually difficult to fulfil
the conditions on the $r_i$ for all the curves $\C_i$.
This will be especially so if there are 'many' contributing curves $\C_i$;
as described in remark b) of sect.~\ref{Pfaff special}, in connection with the
notion of generically K\"ahler-determined bundles,
this is usually the interesting regime.

Note that an exceptional case (for $n\neq 2$) 
implies $\mu\leq \frac{n+1}{2}\chi_{min}-\frac{1}{2}$, 
which gives, for example, for $\chi_{min}=1$ that $\mu=\frac{3}{2}$ for $SU(3)$,
$1$ or $2$ for $SU(4)$ and $\frac{3}{2}$ or $\frac{5}{2}$ for $SU(5)$.

\newpage

\section{\label{Conclusions}Summary and Conlusions}

If one wants to investigate a 'global' string model, 
with gravity not decoupled and not just a string-motivated field theory,
the heterotic string constitutes in many respects still the method of choice
(especially in view of the fact that the precise mathematical 
global geometric study of
singular Calabi-Yau fourfolds, whose singularities are just chosen locally
in the hope that everything fits nicely together, is still in its infancy).
The vast majority of the moduli of such a model is then given by the
moduli of the vector bundle, the relevant superpotential being
the one generated by world-sheet instantons.
Here the conditions $W=DW=0$ for having a supersymmetric Minkowski vacuum
are difficult to evaluate if one is {\em not} in one of the cases where the 
sum over all the contributions from the different instanton curves $\C_i$
vanishes identically anyway: first one has then to handle the phase factor
subtlety to sum up in a reasonable manner the individual contributions,
and second it seems difficult to have control over what might then just be
'accidental cancellations' between the individual contributions.

So an interesting way of producing solutions is to try to make vanish
all the contributions from the different instanton curves {\em individually}
(nota bene: for {\em special} values of the moduli).
Though this may not give all solutions, it may give all 
which are of a reasonable conceptual structure
(given the otherwise accidental nature of the cancellations\footnote{keep 
in mind that we do not speak of cancellations which happen universally
for all moduli - a case for which definitely there are good conceptual
reasons [\ref{BW}] - but only for special moduli}).

The problem with this philosophy is, of course, rooted just in the special
type of such solutions: they demand more conditions - individual 
vanishing (for special values) of the contributions
of all $\C_i$ - than actually necessary (which would
mean that only the sum vanishes). Therefore the following difficulty arises:
even if one finds a set of moduli which makes the contribution, actually 
the Pfaffian $Pfaff_{\C_i}$, of a specific instanton curve $\C_i$ vanish, 
by no means it can be taken for granted that these sets for all the 
different instanton curves $\C_i$ have a non-empty 
intersection $\bigcap_i (Pfaff_{\C_i})$. 
Clearly this should only be the case if the individual conditions 
(making a single Pfaffian vanish) had already some 'global' origin.

It turns out that precisely such a scenario is realised for spectral cover
bundles on elliptically fibered Calabi-Yau spaces $\pi:X\ra B$. However, 
given the specifics of this construction, 
although the (sufficient) vanishing conditions, found in [\ref{C2}] and 
recalled here in Theorem $Pfaff_{single \; \C}$, really fit together
nicely (i.e.~with nonempty intersection)
for a large class of contributions, one has in the end control only over the
contributions from {\em base} instanton curves $\C_i\subset B$, what
constitutes our Theorem $Pfaff_{all \; \C_i\subset B}$ in sect.~\ref{first try}.

Isolating the feature $V|_B\cong \pi_{C*}L$ 
(where $C\ra B$ was a surface cover of the base)
which made progress possible in
this case we define in sect.~\ref{new class} 
a corresponding class of 'projected' bundles where just such a property holds 
over the whole (arbitrary) Calabi-Yau space $X$.
To have control over the different threefold covers $C$ of $X$ we assume 
$C$ embedded in a fourfold $Y$ (not Calabi-Yau).
For this class one can try to adopt the analogy with the scenario of 
spectral bundles over an elliptic $X$ to gain this time control over 
the contributions from {\em all} instanton curves $\C_i\subset X$.
To have as explicit control as possible over the different covers 
$C_3\subset Y_4$ we assume in addition that 
$Y_4$ is fibered over $X_3$; for the purpose of proving our main theorem
we assume that this fibration is elliptic.

So two things have to be done. First one should see 
that this class of bundles satisfies the Donaldson-Uhlenbeck-Yau equation
$F_{a\bar{b}}g^{a\bar{b}}=0$, i.e.~one should make sure that this class
contains examples which have a non-empty region of stability in the 
K\"ahler moduli space $\M_K$. The other thing is, of course, that one 
should carry out the analogy and prove the relevant theorem that there
is a non-empty locus in the moduli space $\M=\M_V\x \M_{cx}$
(actually already in $\M_V$) where the contributions 
from {\em all} Pfaffians $Pfaff_{\C_i}$ vanish. 

We can accomplish both things, though, unfortunately, not at the same time.
We realise that spectral cover bundles on an elliptic Calabi-Yau 
space $X$ constitute an example
of our class of elliptically embedded projected bundles\footnote{note that
the elliptic fibration aspect in the two members of this assertion refers
to different things: first, that $X\ra B$ itself is elliptically fibered, and 
second that the threefold cover $C_3$ lies in an ambient fourfold $Y_4$
which itself is elliptically fibered {\em over} $X$, i.e.~that one has 
a fibration $Y\ra X$}. So, as these spectral bundles are known
to have a non-empty stability region, our class includes indeed stable bundles.

To accomplish our main goal, 
to prove the Theorem $Pfaff_{all \; \C_i\subset X}$
which gives an explicit non-empty locus in the bundle moduli space
where all individual Pfaffian $Pfaff_{\C_i}$ vanish simultaneously,
we need an innocent and canonical looking genericity assumption.
Unfortunately, this assumption is, however, violated in the spectral case.
So, although we reach our main goal to give a theorem which gives
vanishing control over {\em all} instanton contributions $Pfaff_{\C_i}$ 
simultaneously in a class of bundles, the 'elliptically embedded projected'
ones, which contains stable bundles, one still would like to see examples
where genericity and stability can be made sure at the same time.

Apart from this we feel that this strategy, to 
find solutions by making vanish the individual terms in the sum over all $\C_i$
and finding scenarios where the ensuing conditions fit together 'globally',
constitutes a valuable option (if not the only one) to find solutions
of conceptual origin in the vast class of interesting bundles 
where there does {\em not} hold a theorem making the sum vanish
{\em identically}.

I thank B.~Andreas for discussion. I thank
the DFG for support in the project CU 191/1-1 
and the SFB 647 and the FU Berlin for hospitality. 

\appendix

\section{\label{spectral cover appendix}Spectral cover construction}

\resetcounter

Compactification on a Calabi-Yau three-fold $X$ with vector bundle 
$V$ embedded in $E_8\times E_8$ gives a four-dimensional 
heterotic string model of $N=1$ supersymmetry. 
For example, the case of $V$ the tangent bundle 
leads to an unbroken gauge group $E_6$ (times a hidden $E_8$).
The generalisation to an $SU(n)$ bundle $V_1$ gives unbroken GUT
groups like $SO(10)$ and $SU(5)$ (we will in the following focus on
the visible sector and assume a trivial bundle $V_2$ embedded
in the second $E_8$). 

The case where $X$ is elliptically fibered over a base surface $B$
makes possible an explicit description of the bundle by using the
spectral cover surface $C$ of $B$, cf.~sect.~\ref{first try}. 
In this description the $SU(n)$ bundle $V$ is encoded
in two data: a class $\eta\in H^{1,1}(B)$ and a class
$\gamma\in H^{1,1}(C)$ (the latter is connected to the possible
existence of chiral matter in these models [\ref{C}]).
One considers $V$
first over an elliptic fibre $F$ and then globalises over the base $B$. 
Now, over $F$ the bundle $V$ (assumed to be fibrewise semistable)
decomposes as a direct sum of line bundles of degree zero; this is 
described as a set of $n$ points which sum to zero. If one now lets
this vary over the base $B$ it will give a hypersurface 
$C \subset X$
which is a ramified $n$-fold cover of $B$; we assume its class 
given by (with pullback understood)
\beqa
C=n\sigma + \eta
\eeqa
For this one takes $C$ as given as a locus 
$w=a_0+a_2x+a_3y+\dots \, a_nx^{\frac{n}{2}}=0$, 
for $n$ even, say, and $x,y$ the usual elliptic Weierstrass coordinates, 
with $w$ a section of ${\cal O}(\sigma)^n\otimes {\cal N}$ 
(here ${\cal N}$ is a line bundle of class $\eta$);
note that $a_i$ is of class $\eta-ic_1$ (with $c_1:=c_1(B)$).

The idea is then to describe the $SU(n)$ bundle $V$ over $X$ by a line bundle 
$L$ over $C$ 
\beqa
V=p_*(p_C^*L\otimes {\cal P})
\eeqa
with $p:X\times_B C\ra X$ and $p_C: X\times_B C\ra C$ the projections
and ${\cal P}$ the global version of the Poincare line bundle over 
$F\times F$ 
(actually one uses a symmetrized version of this), i.e. the 
universal bundle which realizes the second $F$ in the product as the 
moduli space of degree zero line bundles over the first factor [\ref{FMW}].

A second parameter in the description of $V$ is given by a
half-integral number $\lambda$ which occurs because one gets 
from the condition $c_1(V)=\pi_*(c_1(L)+\frac{c_1(C)-c_1}{2})=0$ 
that with $\gamma \in ker\, \pi_*:H^{1,1}(C)\ra H^{1,1}(B)$ one has
\beqa
\label{line bundle class}
c_1(L)=-\frac{1}{2}(c_1(C)-\pi_{C*}c_1)+\gamma
=\frac{n\si + \eta + c_1}{2}\Big|_C+\gamma
\eeqa
where generally one has for the class $\gamma$ just the following possibility
\beqa
\label{gamma cohomology class}
\gamma=\lambda(n\sigma-\eta+nc_1)|_C
\eeqa
Here $\lambda \in \frac{1}{2}{\bf Z}$ and $\lambda$ 
has to be strictly half-integral for $n$ odd; for $n$ even an integral $\la$
needs $\eta \equiv c_1 \mbox{mod}\, 2$ while a strictly half-integral $\la$
needs $c_1$ even.

Two conditions on $\eta$ are imposed [\ref{OPP}]. First 
the effectivity of $C$ amounts to $\eta$ being effective.
Second to guarantee that $V$ is a stable vector bundle
one assumes irreducibility of $C$
(which follows if the linear system $\eta$ is 
base point\footnote{a base point is a point 
common to all members of the system; on $B$ a Hirzebruch surface ${\bf F_k}$
with base ${\bf P^1}$ $b$ and fibre ${\bf P^1}$ $f$ this amounts to
$\eta \cdot b \geq 0$} free and $\eta - n c_1$ effective).
Often one supposes in addition, as we will do also, 
that $C$ is even ample in $X$
(such that $H^{0,1}(C)=0$ and $L$ is determined by its Chern class); 
then $\eta-nc_1$ is ample in $B$.

For use in sect.~\ref{topological vanishing} 
let us also recall the following facts
\beqa
c_2(X)&=&12c_1\si + 11c_1^2+c_2\\
c_2(V)&=&\eta \si - \frac{n^3-n}{24}c_1^2
+(\la^2-\frac{1}{4})\frac{n}{2}\eta(\eta-nc_1)
\eeqa
Assuming\footnote{to avoid the following restrictions on $\la$ one would have
to use a hidden bundle of non-effective $c_2(V_{hid})$} 
that in the second, hidden $E_8$ a trivial bundle is embedded
one finds for the fivebrane class $W=W_B\si +a_f F$
\beqa
W&=&(12c_1-\eta)\si +11c_1^2+c_2+\frac{n^3-n}{24}c_1^2
-(\la^2-\frac{1}{4})\frac{n}{2}\eta(\eta-nc_1)
\eeqa
Therefore the effectivity demand gives the conditions
\beqa
\eta &\leq & 12 c_1\\
(\la^2-\frac{1}{4})\frac{n}{2}\eta(\eta-nc_1)&\leq &N_B
\eeqa
(where $a\leq b$ for classes means that $b-a$ is effective).
Here we have defined the bundle-independent number $N_B$
(we assume the rank $n$ fixed throughout)
\beqa
N_B&=&(10+\frac{n^3-n}{24})c_1^2+12
\eeqa
using Noether's relation $c_1^2+c_2=12$ for the rational $B$.

Let us now invoke the assumption that $C$ is ample and furthermore that $c_1$
is effective\footnote{which is the case for $B$ 
a Hirzebruch surface ${\bf F_m}$
or a del Pezzo surface ${\bf dP_k}$; 
the latter are more thoroughly described at the end of the next subsection}. 
Then one finds, as $\eta-nc_1$ is ample, 
that $n\leq nc_1(\eta-nc_1)<\eta(\eta-nc_1)$, thus giving
\beqa
n \; < \; \eta (\eta-nc_1)&\leq &\frac{2N_B}{n(\la^2-\frac{1}{4})}
\eeqa
(for $\la\neq \pm 1/2, 0$).
This entails in particular the following restriction on $\la$
\beqa
\label{la bound}
\la^2-\frac{1}{4}&<&\frac{2N_B}{n^2}
\eeqa
For our main example class of bases $B$ with instanton curves, the del Pezzo
surfaces ${\bf dP_k}$, $k=1, \dots, 8$ of $c_1^2=9-k$, one has
$N_{\bf dP_k}\leq N_{\bf F_1}=92+\frac{n^3-n}{3}$.
Thus one has for $SU(3)$ that $\la^2-\frac{1}{4}<200/9$, 
thus leaving (besides $\pm\la=\frac{1}{2}$) the possibilities 
$\pm\la=\frac{3}{2}, \frac{5}{2}, \frac{7}{2}, \frac{9}{2}$.
For $SU(4)$ one gets $\la^2-\frac{1}{4}<14$, 
leaving (besides $\la=0$) the cases $\pm \la=1, 2, 3$.
Finally for $SU(5)$ one has $\la^2-\frac{1}{4}<264/25$, 
thus leaving (besides $\pm\la=\frac{1}{2}$) the possibilities 
$\pm\la=\frac{3}{2}, \frac{5}{2}$.

\subsection{\label{app instanton}
World-sheet instantons contributing to the superpotential}

The (smooth) 
rational curve $\C$ relevant for the world-sheet instanton which we 
consider in sect.~\ref{first try} is a ${\bf P^1}$ lying in the base $B$.
So, besides the $n:1$ covering of $B$ by $C$ which lies in the elliptically
fibered $X$ threefold over $B$, 
one has, with $\E:= \pi^{-1}\C$ the elliptic surface over $\C$
and $c:=C|_{\E}$, the corresponding spectral curve $n:1$ covering
$\pi_c: c\ra \C$ in the elliptic surface $\E$ over $\C$.
Now $C=n\si + \eta$ gives $c=ns+rF$ with the restrictions 
$s:=\si|_{\E}(=\C), r:=\eta \cdot \C$.
Note that $r> n\, c_1\cdot \C=n\, \chi$ (cf.~below) as $\eta-nc_1$ is ample.

Examples are
$B={\bf F_k}$ ($k\leq 2$) where $c_1({\bf F_k})=2b+(2+k)f$ with
$c_1\cdot \C = 2-k$ for $\C=b$ (the base ${\bf P^1}$) 
or $B={\bf P^2}$ of $c_1=3l$ with
$c_1\cdot l=3$ and $c_1\cdot 2l=6$ for $\C$ the line and the quadric, 
respectively. 
Examples on del Pezzo surfaces are discussed below.
Note that
$c _1(\E)=-\pi_{\E}^* \Big(K_{\C}+{\cal O}_{\C}(\chi)\Big)=(2-\chi)F$ with
$\chi:=\chi(\E, {\cal O}_{\E})=\frac{1}{12}e(\E)=c_1\cdot \C$  
(as the discriminant of $Z$ over $B$ is
given by $\Delta = 12 c_1$), so for the mentioned examples one gets
$\E=K3, {\bf dP_9}, b\x F$ for $B={\bf F_k}, \C=b$ with $k=0,1,2$, respectively,
and $e(\E)=36$ or $72$ for the two cases in ${\bf P^2}$.

One has $V|_{B}=\pi_{C*} L$
and so also $V|_{\C}=\pi_{c*}L|_c$. Now [\ref{W}] gives
$W_{\C}\neq 0 \Leftrightarrow V|_{\C}$ trivial or
\beqa
W_{\C}\neq 0 \Llra
0=h^0\Big(\C, V|_{\C}\ox\cO_{\C}(-1)\Big)= 
h^0\Big(c, L|_c\ox \cO_c(-F|_c) \Big)
\eeqa
(we will use also the notation $V(-1)|_{\C}:=V|_{\C}\ox\cO_{\C}(-1)$
and $L(-F)|_c:=L|_c\ox \cO_c(-F|_c)$)
where one has explicitly
\beqa
\label{concrete line bundle}
L|_c\ox \cO_c(-F|_c)=\cO_{\E}\Bigg(n \Big(\la + \frac{1}{2}\Big) s 
+ \Big[\Big(n\la + \frac{1}{2}\Big)\chi-\Big(\la - \frac{1}{2}\Big)r-1\Big] 
F\Bigg)\Bigg|_c
\eeqa

Now recall that $N_X \C = \cO_{\C}(a)\oplus \cO_{\C}(-a-2)$ where, say,
$a=\C^2$, the selfintersection number in $B$, and $-a-2=s^2$,
the selfintersection number of $\C$, now considered as (the base) 
curve $s$ in $\E$; note that one has in $\E$ (when suitably interpreted)
$-2=K_s=s^2+K_{\E}\cdot s=s^2+\chi -2$, that is $\chi=-s^2$.

If we now take also into account that we want to assume that $\C$ is 
(actually even infinitesimally) 
isolated one gets $N_X \C = \cO_{\C}(-1)\oplus \cO_{\C}(-1)$ 
such that\footnote{note that $\chi=\C\cdot c_1\not\equiv 0\, \mbox{mod}\, 2$
excludes the case $c_1$ even 
and so also the case $n$ even with $\la\in \frac{1}{2}+{\bf Z}$}
$\chi=-s^2=1$. Therefore among the mentioned examples just the base $b$
in ${\bf F_1}$ remains. This is just the del Pezzo surface ${\bf dP_1}$.
So, as we search in particular for (smooth) rational $(-1)$-curves
we have now a closer look on a del Pezzo surface as base 
and its $(-1)$-curves.\\

{\em The del Pezzo surfaces ${\bf dP_k}$}

We collect some facts about del Pezzo surfaces
which are examples to be used as base surfaces $B$ of elliptic
Calabi-Yau threefolds. Their interest in our set-up lies in the fact
that they come with many smooth isolated rational curves, often called
$(-1)$ - curves from their self-intersection; we will call them instanton
curves.

The del Pezzo surface\footnote{sometimes the rational
elliptic surface $"{\bf dP_9}"$ is included which comes with infinitely many
$(-1)$ - curves; for other reasons sometimes the Hirzebruch surface 
${\bf F_0}={\bf P^1}\x {\bf P^1}$ is included; note that ${\bf dP_0}={\bf P^2},
{\bf dP_1}\cong {\bf F_1}, {\bf dP_5}\cong {\bf P^4}(2,2), 
{\bf dP_6}\cong {\bf P^3}(3),
{\bf dP_7}\cong {\bf P_{1,1,1,2}}(4), {\bf dP_8}\cong {\bf P_{1,1,2,3}}(6)$} 
${\bf dP_k}$ for $k=0, \dots , 8$
is the blow-up of ${\bf P^2}$ at $k$ points (lying suitably general, i.e.~no
three points lie on a line, no six on a conic).
The exceptional curves from these blow-ups are denoted by $E_i, i=1, \dots, k$.
The intersection matrix for $H^{1,1}({\bf dP_k})$ 
in the basis $(l, E_1, \dots, E_k)$
with the proper transform $l$ of the line from ${\bf P^2}$ is just
$Diag(1, -1, \dots, -1)$; 
furthermore $c_1({\bf dP_k})=3l-\sum_i E_i$ such that $c_1^2({\bf dP_k})=9-k$.

There are for $k\geq 2$ 
further instanton curves on the ${\bf dP_k}$ besides the $E_i$ themselves.
First one has the 
$\footnotesize{\Big(\!\begin{array}{c}k\\2\end{array}\!\Big)}$
proper transforms of lines in ${\bf P^2}$ through two different blow up points;
they have in $dP_k$ the classes $l-E_i-E_j$.
Then there are in addition for the higher ${\bf dP_k}$ of $k\geq 5$ some 
exceptional configurations: 
there are 
$\footnotesize{\Big(\!\begin{array}{c}k\\5\end{array}\!\Big)}$
curves (proper transforms of conics through five of the points)
of classes $2l-\sum_{n=1}^5 E_{j_n}$ 
(all occurring $E$-indices $j_n$ (and $i$) from the set $1, \dots, k$ 
have to be different; similarly for the cases below);
for $k \geq 7$ there are
$k \cdot \footnotesize{\Big(\!\begin{array}{c}k-1\\6\end{array}\!\Big)}$
curves of classes $3l-2E_i-\sum_{n=1}^6 E_{j_n}$
(these are proper transforms of singular cubics through all the seven points,
with a double point at one of them);
finally on ${\bf dP_8}$ there are even more special curves:
$56$ curves of class 
$4l-2\sum_{n=1}^3 E_{j_n}-\sum_{m=1}^5 E_{j_m}$, 
$28$ curves of class $5l-2\sum_{n=1}^6 E_{j_n}-\sum_{m=1}^2 E_{j_m}$
and $8$ curves of class $6l-3E_i -2\sum_{n=1}^7 E_{j_n}$.

\subsection{\label{technical}A technical aside on the spectral cover equation}

Before we come in the app.~\ref{proof appendix} to the proof of the fact
that a special property of the coefficients of the spectral cover equation
leads to zeroes of the Pfaffian we take - for convenience of the reader - 
the opportunity to make clear a technical point on this equation.
The spectral surface $C$ is given by (\ref{C equation})
with $w\in H^0(X, {\cal O}(\sigma)^n \ox \pi^*\N)$ where
\beqa
\label{explicit coordinate forms}
\un{n=2}& \hspace{1.5cm}& w\, = \, a_0z+a_2 x\nonumber\\
\un{n=3}& \hspace{1.5cm}& w\, = \, a_0z+a_2 x+a_3 y\nonumber\\
\un{n=4}& \hspace{1.5cm}& w\, = \, a_0z^2 + a_2 xz + a_3 yz + a_4 x^2 
\nonumber\\
\un{n=5}& \hspace{1.5cm}& w\, = \, a_0z^2 + a_2 xz + a_3 yz + a_4 x^2 + a_5 xy
\nonumber\\
\un{n=6}& \hspace{1.5cm}& w\, = \, a_0z^3 + a_2 xz^2 + a_3 yz^2 + a_4 x^2z 
+ a_5 xyz + a_6 x^3
\eeqa
or in general (where $0\leq i, j$ and $j\leq 1$; $w$ has degree 
$[\frac{n}{2}]=
\left\{ \begin{array}{ll}
\frac{n}{2}\;\;\;\;\; n \, \mbox{even}\\
\frac{n-1}{2}\;\; n \, \mbox{odd}
\end{array} \right. $ in $x,y,z$)
\beqa
w & = & \sum_{\stackrel{\scriptstyle m=0}{2i+3j=m\neq 1}}^n
a_m x^i y^j z^{[\frac{n}{2}]-(i+j)}\\
&=&a_0 z^{[\frac{n}{2}]} + a_2 x z^{[\frac{n}{2}]-1} 
+ a_3 yz^{[\frac{n}{2}]-1} + \dots \nonumber\\
&&
\;\;+\;\left\{ \begin{array}{ll}
a_{n-2}x^{[\frac{n}{2}]-1}z + a_{n-1}x^{[\frac{n}{2}]-2}yz \, + \;
a_n \, x^{[\frac{n}{2}]}\;\;\;\;\;\;\;\;\;\;\;\;\;\;\;\;\;\;\;\;\;\;\;
n \, \mbox{even}\\
a_{n-3}x^{[\frac{n}{2}]-1}z   + 
a_{n-2}x^{[\frac{n}{2}]-2}yz \, + a_{n-1}x^{[\frac{n}{2}]}
+a_n x^{[\frac{n}{2}]-1}y \;\;\; n \, \mbox{odd}
\end{array} \right. 
\nonumber
\eeqa
Now let us consider the situation on a single fibre $F$.
For the elliptic curve $F\subset {\bf P^2_{x,y,z}}$ 
given by the Weierstrass equation $zy^2=4x^3-g_2xz^2-g_3z^3$
(with zero point $p_0=(0,1,0)$), 
the divisor $(z)=l\subset {\bf P^2}$ becomes $(z)|_F=3p_0$ on $F$. 

\noindent
\un{\em $n$ even}
To encode $n$ points on $F$ one 
chooses a homogeneous polynomial $w^{(hom)}_{n/2} (x,y,z)$ of degree $n/2$. 
From its $3 n/2$ zeroes on $F$ only $n$, say $q_i$,
carry information as $n/2$ of them are always at $p_0$: the rewriting
$w^{(hom)}_{n/2}(x,y,z)=z^{n/2}w^{aff}_{n/2}(x/z,y/z)$ shows 
$3 n/2$ zeroes at $p_0$ from the $z$-power, 
and $n$ poles at $p_0$ and $n$ zeroes
at the $q_i$ from the 
affine factor.\footnote{$x/z$ and $y/z$ have a pole at $p_0$ of order
$-(1-3)=2$ and $-(0-3)=3$, resp.} 
So one gets on $F$ for the divisor of $w^{(hom)}$
resp. for the divisor of zeroes of $w^{aff}$
\beqa
\label{subtlety even}
(w^{(hom)}|_F)=\frac{n}{2}p_0 + \sum^n q_i\;\;\;\; , \;\;\;\;
\;\;\;\; (w^{aff}|_F)_0=\sum^n q_i
\eeqa

\noindent
\un{\em $n$ odd}
Similarly here extracting a factor $z^{\frac{n-1}{2}}$ gives
(when $n$ poles at $p_0$ of the affine part cancel $n$ zeroes of
the $3\cdot \frac{n-1}{2}$ zeroes of $z^{\frac{n-1}{2}}|_F$)
\beqa
\label{subtlety odd}
(w^{(hom)}|_F)=\frac{n-3}{2}p_0 + \sum^n q_i\;\;\;\; , \;\;\;\;
\;\;\;\; (w^{aff}|_F)_0=\sum^n q_i
\eeqa

Now two cases deserve special attention here.
First one may ask: when is at least 
one of the $n$ zeroes $q_i$ the point $p_0$ ?
Considered in $X$ (or in $\E_i$) this is, of course, the question
at which points $C$ intersects $B$ (or $c_i$ intersects $\C_i$).
The answer, involving the zero divisor $(a_n)$, 
was given (in a reasoning in affine coordinates) in equ.~(6.8) in [\ref{FMW}]
\beqa
\label{zeroes in the base}
B\cap C &=& (a_n)\;\;\;\;\;\;\;\;\;\;\;\; 
(\;\mbox{or}\;\; \C_i\cap c_i\, = \, (a_{n;i})\; )
\eeqa

Of course, this is also seen in the factorization of the 
homogeneous polynomial above: if one is
at a point $b\in B$ (or $b\in \C_i$) where the highest coefficient of 
the spectral cover equation vanishes,
then the affine factor $w^{aff}|_{F_b}$ has only $n-1$ zeroes $q_i$ and
$n-1$ poles at $p_0$ such that 
one zero less of the $z$-power $z^{[\frac{n}{2}]}|_{F_b}$ 
is cancelled, giving the net effect
that one of the previous (generic) $n$ zeroes $q_i$ is now at $p_0$.
Conversely if one of the $q_i$ is at $p_0$ it cancels one of the poles there;
but as the number of poles is manifest in the occurring highest power
this means that its coefficient vanishes.

A second case is decisive for the theorems in the present paper, 
cf.~app.~\ref{proof appendix}.
This is the case when now not just 
{\em one} of the $n$ points $q_i$ on a fibre
(encoded by the spectral cover equation) lies actually at $p_0$ but when
{\em all} $n$ zeroes $q_i$ are at $p_0$.
From what was said a moment ago in the first case it is clear that
this will just happen if all $n$ poles of the affine factor drop out.
This in turn will just happen if all coefficients higher than $a_0$ vanish.
As we have seen, that\footnote{now we describe the situation in the set-up
$c_i\subset \E_i$ over $\C_i$} at all points $u_{j; i}$
of $(a_n)\cap \C_i=
(a_{n;i})\subset \C_i$ already one of the spectral points (the zeroes $q_i$)
is at $p_0$, the condition that {\em at these points} $u_{j; i}$
the coefficients $a_{2;i}, \dots, a_{n-1;i}$ vanish as well
is just given by $a_{n;i} | a_{j;i}$ for all $j=2, \dots, n-1$.

This is just the relation $\R_i\subset \un{\Sigma_i}$ (actually $"="$)
in app.~\ref{Characterisation} which we recall in 
the proof of assertion a) of Theorem $Pfaff_{single\; \C\subset B}$ in 
sect.~\ref{proof appendix} below: along this locus $\R_i$ 
(of expected codimension $(n-2)(r_i-n)$)
in moduli space
$\M_{\E_i}(c_i)$ the $n$ fibre points of $c_i$ over each of the 
zeroes $u_{j;i}$  of $a_{n;i}$ on $\C_i\cong {\bf P^1}$
will {\em all} be just $p_0$ because now not only $a_{n;i}$ vanishes
at these points of $\C_i$ but actually all the $a_{n;j}$, $j=2, \dots, n-1$,
as well.

\un{\em Remark:} One realises $(r_i-n)(n-2)$ 
(cf.~the interpretation of this number in app.~\ref{conceptual}) 
as the codimension of this
specialisation locus in $\M_{\E_i}(c_i)$ 
by computing for $h^0(\E_i, \cO_{\E_i}(c_i))$
\beqa
h^0\Big(\C_i, \cO_{\C_i}(r_i)\Big)
+\sum_{j=2}^{n-1}h^0\Big(\C_i, \cO_{\C_i}(r_i-j)\Big)
+h^0\Big(\C_i, \cO_{\C_i}(r_i-n)\Big)
&=&nr_i-\frac{n^2-n}{2}\;\;\;\;\;\;\;\;\;\;\;\;
\eeqa
whereas\footnote{cf.~with regard to this comparison also the 
remark a) in sect.~\ref{concrete solution set}} 
one gets for the degrees of freedom in the specialisation $\R_i$
\beqa
h^0\Big(\C_i, \cO_{\C_i}(r_i)\Big)
+\sum_{j=2}^{n-1}h^0\Big(\C_i, \cO_{\C_i}(n-j)\Big)
+h^0\Big(\C_i, \cO_{\C_i}(r_i-n)\Big)
&=&2r_i+\frac{n^2-3n}{2}\;\;\;\;\;\;\;\;\;\;\;\;
\eeqa

\subsection{\label{Characterisation}Characterisation of the solution locus}

Let us try to understand more fully the conditions in the sets $\R$ and $\Si$,
or their respective local versions $\R_i$ and $\Si_i$ over an 
individual instanton curve $\C_i$ in $B$, 
and along this way provide a proof to the proposition 
stated in sect.~(\ref{concrete solution set}).
First one realises that here naturally 
a further locus $\un{\Sigma}$ in the moduli space occurs 
with the following 'logical position'
\beqa
\R \;\; \subset \;\; \un{\Sigma} \;\; \subset \;\; \Sigma 
\eeqa
For this let us first define a corresponding locus $\un{\Sigma_i}$ 
in the situation for the Pfaffian for one individual instanton curve
(we use the zeroes $u_{k;i}$ of $a_{n;i}$ where $k$ runs from $1$ to $r_i-n$;
$s_i|_{c_i}$ is in each of the fibers just the respective point $p_0$; 
here $\sim$ is linear equivalence)
\beqa
\label{def unSigma_i}
\un{\Sigma_i}&:=&\Big\{ t\in \M_{\E_i}(c_i)\Big| \; \;
ns_i|_{c_i}\, = \, \sum_{k=1}^{r_i-n}\, F_{u_{k;i}}|_{c_i}\Big\}\\
\Sigma_i&=&\Big\{ t\in \M_{\E_i}(c_i)\Big| \; \;
ns_i|_{c_i}\, \sim \, (r_i-n)F|_{c_i}\Big\}
\eeqa
(here for comparison we also recalled an explicit form of the condition
defining $\Sigma_i$).
Now we enlarge the consideration from one instanton curve $\C_i$
to $B$ as a whole and define correspondingly,
in analogy to (\ref{def unSigma_i}), 
using the curve $A:=\sigma\cap C=(a_n)$,
\beqa
\label{def unSigma}
\un{\Sigma}&:=&\Big\{ t\in \M_X(C)\Big| \; \;
n\sigma|_C\; = \; \pi^{-1}(A)|_C\Big\}
\eeqa

Now the strategy for proving the mentioned proposition of 
sect.~(\ref{concrete solution set}) consists in two parts.
First one realises that one has directly the inclusions
$\R_i\subset \un{\Si_i}\subset \Si_i$, respectively 
$\R\subset \un{\Si}\subset \Si$. Then we will show the opposite inclusions;
actually we proceed by treating these reasonings first for the 
first inclusion (showing that it is actually an equality) and 
then, in a more extended argument, for the second inclusion
(although all of this is completely elementary we give the arguments in detail).

So note first that not just the obvious 
(cf.~app.~\ref{technical} and app.~\ref{proof appendix}) relations 
$\R_i\subset \un{\Si_i}$ and $\R\subset \un{\Si}$ hold
but that one actually has $\R_i=\un{\Si_i}$ and $\R=\un{\Si}$.
The first equality is explained in app.~\ref{technical}. The global version
$\R=\un{\Si}$
follows from the same reasoning as the property of a $C$-fibre $F_b|_C$ 
(over a point $b\in B$) 
to have all its $n$ points at $p_0=F_b\cap \sigma$
is shown in app.~\ref{technical} to be equivalent
to the vanishing of $a_j(b)$ for $j=2, \dots, n$.
Only the relation $\Sigma \subset \R$ remains to be seen.
But the condition $a_j|_A=0$ for $j=2, \dots, n-1$ indeed entails
$a_j=a_j'\cdot a_n$ as the chain of $H^0$-terms in the long exact sequence 
associated to the restriction to $A$ (of the respective bundle of which 
an $a_j$ is a section) shows.

Note further that also not just the obvious relations 
$\un{\Si_i}\subset \Si_i$ and $\un{\Si}\subset\Si $ hold
but that one actually has $\un{\Si_i}=\Si_i$ and $\un{\Si}=\Si$.
Though elementary let us give the reasoning in detail.
As $ns_i|_{c_i}$ is effective there exists a nonzero section
$f\in H^0(c, \cO_{\E_i}(ns_i)|_{c_i})$ which fulfils of course, as {\em any}
such section, the relation $(f)\sim ns_i|_{c_i}$.
On the other hand {\em there exists} a specific meromorphic section $g$
of the bundle in question with $(g)=ns_i|_{c_i}$; as the latter divisor is
effective $g$ is actually a holomorphic section, i.e.~an element of $H^0$;
so we can take an $f$ as before with 
\beqa
(f)&=&ns_i|_{c_i}
\eeqa

Now, for a modulus $t\in \Si_i$ for which $\cO_{\E_i}(ns_i-(r_i-n)F)|_{c_i}$
becomes trivial there exists a nonzero holomorphic section $\zeta$ of this
bundle on the curve $c_i$
(as the line bundle in question is flat the existence of such a $\zeta$
is even equivalent for the bundle being trivial: the 
linearly equivalent {\em effective} divisor of $\zeta$ also has
degree zero and thus is zero). Clearly, as the line bundle is trivial,
$\zeta$ has neither zeroes (nor poles). In other words the element
$f/\zeta$ is a nonzero section of $\cO_{\E_i}((r_i-n)F)|_{c_i}$
\beqa
\frac{f}{\zeta}&\in&H^0\Bigg(c_i, \cO_{\E_i}\Big((r_i-n)F\Big)\Big|_{c_i}\Bigg)
\eeqa
which still has divisor 
\beqa
\label{section representation}
\Bigg(\frac{f}{\zeta}\Bigg)&=&ns_i|_{c_i}
\eeqa

At this point a priori one knows only that 
$(\frac{f}{\zeta})\sim (r_i-n)F|_{c_i}$. But it turns out that now
one has actually (where the $b_k$ must be the points of $s_i\cap c_i$)
\beqa
\label{concrete fibre representation}
\Bigg(\frac{f}{\zeta}\Bigg)&=& \sum_{k=1}^{r_i-n}F_{b_k}|_{c_i}
\eeqa
For this one may note that any relation $D\sim F|_{c_i}$ entails
$D=F_{b_*}|_{c_i}$ as the linear system $\Big| F|_{c_i} \Big|$ is 
complete, i.e.~contains any effective divisor which is linearly equivalent 
to any of its members and these members turn out to be just all the
$F_{b}|_{c_i}$ (the corresponding property holds of course for the linear
system $|F|$ on $\E_i$). This is easily inductively generalised for the
corresponding relevant property for a multiple fibre class, thereby proving 
(\ref{concrete fibre representation}).\footnote{An alternative reasoning 
might use the fibration structure instead as follows: from the long exact
sequence associated to the short exact sequence involving the restriction
from $\E_i$ to $c_i$ for the bundle $\cO_{\E_i}((r_i-n)F)$ one finds
that $H^0(c_i, \cO_{\E_i}((r_i-n)F)|_{c_i})\cong 
H^0(\E_i, \cO_{\E_i}((r_i-n)F))$, the latter being in turn equal to
$\pi^*H^0(\C_i, \cO_{\C_i}(r_i-n))$; this realises $f/\zeta$ as a 
pullback polynomial $\pi^*P_{r_i-n}$ restricted to $c_i$ which itself
has the sought-after divisor.} Then (\ref{section representation}) and 
(\ref{concrete fibre representation}) prove that $\Si_i\subset \un{\Si_i}$.\\
\indent
The argument for the global case $\Si\subset \un{\Si}$ is similar: 
one finds the divisor $(f/\zeta)=n\sigma|_C$ 
which is $\sim \pi^{-1}A$ and is actually $=\pi^{-1}A'$ with $A'\in |A|$
where the $A'$ must be $\sigma\cap C=A$.

Thus in total we have by now shown that one has
\beqa
\R_i \; = \; \un{\Si_i}\; = \; \Si_i\\
\R \; = \; \un{\Si}\; = \; \Si
\eeqa
Thereby one has proven the statements a) and b) of the proposition 
in sect.~\ref{concrete solution set}.

\un{\em Remark:}
This may be the right place to point yet to a further interpretation
of the specialisation condition which defines our solution locus
$\Si_i$ or $\Si$ in moduli space. It is related directly to the 
degenerate structure for the spectral cover $C$ which the conditions impose.\\
\indent
As along $\un{\Si}$ (in moduli space) the behaviour of $C$, or $V$,
along $A$ (in $X$)
is decisive let us describe the effect of the spectral degeneration
there further (for more on this cf.~sect.~\ref{location subsection}).
The right hand side in (\ref{def unSigma_i}) can be written also as follows:
the locus on $s_i$ given by the zero divisor of $a_{n;i}$ is just $s_i\cap c_i$,
and $c_i$ is everywhere over $s_i$ just an $n$-fold collection of points in 
the fibres $F$ over $s_i$; so one can write as characterising condition
for the set $\un{\Sigma_i}$ equivalently
\beqa
\label{def unSigma_i alternativ}
\un{\Sigma_i}&=&\Big\{ t\in \M_{\E_i}(c_i)\Big| \; \;
ns_i|_{s_i\cap c_i}\; = \; c_i|_{\pi^{-1}(s_i\cap c_i)}\Big\}
\eeqa
This is in the situation for one instanton curve in the base.
Now going back to $B$ as a whole one has correspondingly,
in analogy to (\ref{def unSigma_i alternativ}),\footnote{in 
(\ref{def unSigma_i alternativ}) one may write
$n(s_i\cap c_i)$ for $ns_i|_{s_i\cap c_i}$, 
and in (\ref{def unSigma}) similarly $nA$ for $n\sigma|_A$}
\beqa
\un{\Sigma}&=&
\Big\{ t\in \M_X(C)\Big| \; \;
n\sigma|_A\; = \; C|_{\pi^{-1}(A)}\Big\}\;\;\;\;\;\;\;\;\;
\eeqa
Now, an $SU(m)$ bundle on $X$ which arises
as pullback $\pi^*M$ of an $SU(m)$ bundle $M$ on the base $B$ has the degenerate
spectral data $(m\sigma, M)$, i.e.~its spectral cover surface $C$ is the
nonreduced object given by taking $B$ with multiplicity $m$. Although we 
expect a line bundle on the spectral cover surface and not a rank $m$ bundle,
here the rank $m$ bundle $M$ (which splits, of course, 
as a direct sum of $m$ one-dimensional spaces over each point of $B$) 
is understood as a deformation of a line bundle over $m\sigma$.\\
\indent
Now we see that - along the locus $\un{\Sigma}$ in moduli space - the part
$C|_{\pi^{-1}(A)}$ of the spectral cover surface $C$ which lies over $A$ 
has according to (\ref{def unSigma}) just such a nonreduced structure.
So, over this subset $A$ the role of $M$ is played here by
$L|_A\oplus L|_A\oplus \dots \oplus L|_A = \oplus_1^n L|_A$. 
So, from the previous paragraph, $V$
has now along $A$ also the further\footnote{note that in any case $V$ 
has along $B= \sigma$ 
the representation $V|_B\cong \pi_{C*}L$ and has therefore along $A$
the representation $V|_A\cong \pi_{(C\cap \pi^{-1}(A))*}L|_{\pi^{-1}(A)}$}
representation $V|_A\cong (\pi^* L|_A)|_{nA}$, or rather 
one gets that along $\un{\Sigma}$ one has 
\beqa
V|_A&\cong & \oplus^n L|_A
\eeqa
This is another manifestation of the degenerate structure the conditions impose.

\subsection{\label{conceptual}Conceptual argument for the main implication}

The source from which everything flows in this note is
Theorem $Pfaff_{single \, \C\subset B}$, more precisely the 
principle of its proof. This is generalised to include the cases
of {\em all} base curves for a spectral bundle 
on an elliptic $X$
and then in sect.~\ref{new class} to the case of certain projected bundles
over an arbitrary $X$. We give the proof of this basic assertion 
in app.~\ref{proof appendix} and did recall the idea already 
in app.\ref{technical}. The reasoning relies 
on the use of the concrete spectral equation for the cover object
(the curve $c_i\subset \E_i$ over $\C_i$).
As this explicit description may be not available in related set-ups 
it clearly would be of interest to have a more conceptual understanding
of the argument which is completely independent of the use of a concrete
defining equation of the cover object. Here we will keep the concrete structure
of our case but {\em sketch} 
an argument not using the equation, and so in particular 
also not using the representation of the moduli by the coefficients polynomials
$a_{n;j}$.

So let us consider in the elliptic surface $\E_i$ the cover curve $c_i$
of class $ns_i+r_i F$ together with the relevant line bundle
$\cO_{\E_i}(ns_i-(r_i-n)F)$. Its restriction to $c_i$ has degree zero
and the issue at stake is that the part $-(r_i-n)F$ of the divisor violates
manifestly the effectivity on $\E_i$. One wants now 
to show that a concrete curve
$c_i$, i.e.~a member in the linear system $|c_i|$ of curves in $\E_i$,
exists for which $ns_i|_{c_i}$ and $(r_i-n)F|_{c_i}$ are linearly equivalent.
We show that a member exists for which these divisors on $c_i$ are even equal.

First (concerning the zeroes $u_j\in \C_i=s_i$ of $a_{n;i}$)
note that there exist $r_i-n$ concrete fibers $F_j|_{c_i}$, 
i.e.~members of the linear system $|F|_{c_i}|$, which contain
the zero point $p_0$ of the respective fibre (as $s_i|_{c_i}$
has $s_i\cdot c_i=r_i-n$ points $u_j$ and $|F|_{c_i}|$ covers the whole $c_i$).

Secondly (replacing the argument involving the condition
$a_{n;i}|a_{j;i}$) we ask for the existence of a concrete $c_i$ (a member of
$|c_i|$) with $\sum_{j=1}^{r_i-n}F_j|_{c_i}=ns_i|_{c_i}$ 
(the sum taken over the concrete fibers of the first step)\footnote{note that if
one of these fibers contains in addition to $p_0$ further 
$n-2$ times the point $p_0$ then already {\em all} $n$ fibre points are $p_0$
as they have to sum up to zero (represented by $p_0$) 
in the group law on the fibre}; to 
specialise $n-2$ points to $p_0$ in each of the $r_i-n$ fibers amounts to
$(r_i-n)(n-2)$ conditions\footnote{it will then turn out that each of these 
counts just as one codimension (cf.~the remark in app.~\ref{technical})}.
Now the 'restriction' map\footnote{think of the evaluation 
(\ref{restriction evaluation}) as giving the value 
and the first $k$ derivatives (at the $r_i-n$ points $u_j$ of $s_i|_{c_i}$)
of a 'function' with zero set $c_i$; clearly all the values 
($0$'th order derivatives) vanish already}  
(for $k\geq 0$)
\beqa
\label{restriction evaluation}
H^0\Big(\E_i, \cO_{\E_i}(c_i)\Big)&\ra& 
\bigoplus_{u_j\in s_i|_{c_i}}
H^0\Big( u_j, \cO_{u_j}/I^{(u_j)}_{k+1}\Big)
\; \cong \; \bigoplus^{r_i-n}\; \bigoplus^{k+1}\, {\bf C}\;\;\;\;\;\;\;\;
\eeqa
(after shown to be surjective)
has a kernel of dimension
$nr_i-\frac{n^2-n}{2}+1-k(r_i-n)$ which is ($1+$) the
dimension of the linear subsystem of $|c_i|$ of curves 
containing $s_i|_{c_i}$ even with multiplicity $k+1$ 
(and not just the automatic multiplicity $1$); 
$k=n-2$ gives the assertion.

\newpage

\section{\label{proof appendix}
Proof of Theorem $Pfaff_{single\, \C\subset B}$}

\resetcounter

We recall the proof of parts a) and c)
of Theorem $Pfaff_{single\, \C\subset B}$ 
from sect.~\ref{concrete solution set} 
(for the first step here, the proof of a), cf.~also 
the reasoning in app.~\ref{technical} and \ref{conceptual}). 

a) \underline{$\R_i\subset \Sigma_i$:}
In view of the equation\footnote{written in affine 
coordinates $(x,y)$ of the ${\bf P^2}$ fibre} (\ref{C equation}) for $C$,
or the corresponding ensuing 
equation\footnote{for $n$ even, say; now written in projective 
coordinates $(x,y,z)$; 
note the subtleties (\ref{subtlety even}),(\ref{subtlety odd})} 
$w_i:=a_{0;i}z^{n/2}+a_{2;i}xz^{n/2 -1}+a_{3;i}yz^{n/2 -1}
+\dots +a_{n;i}x^{n/2}=0$,
one finds along the locus $\R_i$ in moduli space a special coincidence of
two in general different divisors on $c_i$: first one has the divisor
$s_i|_{c_i}$ which consists of $s_i \cdot (ns_i+r_i F)=r_i-n$ points 
(counted with multiplicities) which, 
according to equ.~(\ref{zeroes in the base}), 
lie in fibers $F_{u_k}|_{c_i}$ of $c_i$ over the 
$\mbox{deg}\, a_{n;i}=r_i-n$ zeroes $u_k$
of the homogeneous polynomial $a_{n;i}$ on the ${\bf P^1}$
given by\footnote{$s_i$ is the base (and section) of the elliptic surface
$\E_i:=\pi^{-1}(\C_i)$ which shares the general elliptic fibre $F$ with $X$;
that is, $s_i$ (which is also $\sigma\cap \E_i$) 
is the curve $\C_i$ considered in $\E_i$ instead of $B$} 
$s_i$. On the other hand clearly
any of these fibers will consist of $n$ points of $c_i$. It would be a rather
special circumstance if the corresponding $n$ points in each of the $r_i-n$
fibers $F_{u_k}|_{c_i}$ 
over the zeroes of $a_{n;i}$ would just consist of the intersection
point (of coordinates $(x,y,z)=(0,1,0)$ in the ${\bf P^2}$ coordinates)
of $c_i$ with the section $s_i$ {\em counted $n$ times}.
But this is just what happens along the locus $\R_i$ as there
(cf.~sect.~\ref{technical}), to be in
one of the fibers $F_{u_k}$, implies $z=0$, i.e. to be on the section $s_i$.
Therefore the divisors $ns_i|_{c_i}$ and $\sum_{k=1}^{r_i-n}F_{u_k}|_{c_i}
\sim (r_i-n)F|_{c_i}$ (where $\sim$ is linear equivalence)
are the same, and so the line bundle on $c_i$ corresponding to the difference
of these divisors is trivial indeed.

\underline{$\Sigma_i\subset Pfaff_{\C_i}$:}
Because of $h^0(\C_i, V|_{\C_i}\ox \cO_{\C_i}(-1))=
h^0(c_i, L|_{c_i}\ox \cO_{c_i}(-F|_{c_i}))$ one needs to find 
a nontrivial section of $L|_{c_i}\ox \cO_{c_i}(-F|_{c_i})$.
Now, concretely one has with (\ref{concrete line bundle})
\beqa
L|_{c_i}\ox \cO_{c_i}(-F|_{c_i})&=&\left\{ \begin{array}{ll}
\cO_{\E_i}\Big(\frac{1}{2}\Big[ns_i+(r_i-1)F\Big]
+\la\Big[ns_i-(r_i-n)F\Big]\Big)\Big|_{c_i}
& \mbox{for} \; n \, \mbox{even} \\
\cO_{\E_i}\Big((\la+\frac{1}{2})ns_i
+\Big[(\la n - \frac{1}{2})-(\la-\frac{1}{2})r_i\Big]F\Big)\Big|_{c_i}
& \mbox{for} \; n \, \mbox{odd}\end{array} \right.\nonumber\\
\label{specialisation}
&\stackrel{\Sigma_i}{\ra}&\left\{ \begin{array}{ll}
\cO_{\E_i}\Big(\frac{n}{2}s_i+\frac{r_i-1}{2}F\Big)\Big|_{c_i}
& \\
\cO_{\E_i}\Big(\Big[r_i-\frac{n+1}{2}\Big]F\Big)\Big|_{c_i}
& 
\end{array} \right.
\eeqa
(to use $ns_i|_{c_i}\sim (r_i-n)F|_{c_i}$
we write first different forms of the same expression 
because of the half-integrality issue; 
note that for $n$ even one has $\la \in {\bf Z}$ and $r_i$ odd, while
for $n$ odd one has $\la \in \frac{1}{2}+{\bf Z}$). 
(\ref{specialisation}) shows the effectivity
of the relevant divisors along $\Sigma_i$.

c) First one gets, in refinement of the criterion 
$Pfaff_{\C}(t)=0 \Longleftrightarrow
h^0(c_t, L(-F)|_{c_t})\neq 0$, that the vanishing order of $Pfaff_{\C}(t)$
is bounded below by $h^0(c_t, L(-F)|_{c_t})$ (cf.~(4.38) in [\ref{C2}]).
From the long exact sequence one finds that the sections of the corresponding
bundles on $\E$ in (\ref{specialisation}) {\em inject} 
into the sections of the bundles restricted to $c_t$.
This gives the assertion when using (for $n$ even; note $r>n$) that
$h^0(\E, \cO_{\E}(\frac{n}{2}s+\frac{r-1}{2}F))
=h^0(\C, \cO_{\C}(\frac{r-1}{2})
\oplus \bigoplus_{i=2}^{n/2}\cO_{\C}(\frac{r-1}{2}-i))
=\frac{r-1}{2}+1+\sum_{i=2}^{n/2}(\frac{r-1}{2}-i+1)=\frac{n}{2}(\frac{r-1}{2}
+1)-(\frac{1}{2}\frac{n}{2}(\frac{n}{2}+1)-1)$.

\section{\label{deviations}Deviations in the proof of theorem
$Pfaff_{all\, \C\subset X}$}

\resetcounter

The essential deviation in the set-up where we have an elliptically fibered
fourfold $Y$ (not being a Calabi-Yau space) over our arbitrary Calabi-Yau
threefold $X$ lies in the fact that now no longer the isolatedness of an
instanton curve $\C_i$ in $X$ is related to the value of $\chi_i:=\C_i\cdot \la$
where $\la=c_1(\cL)\in H^2(X)$ encodes the fibration 
(cf.~(\ref{la definition}); 
previously this was the class $c_1=c_1(B)\in H^2(B)$, and we found $\chi_i=1$
in app.~\ref{app instanton}).

Having now many individual $\chi_i$ (which is still $-s_i^2$, 
but $\E_i=\pi^{-1}(\C_i)$ will be no longer ${\bf dP_9}$ in general), 
let us see
how the arguments in the proof of Theorem $Pfaff_{single\, \C\subset B}$
have to be adapted.
{\em We assume $\la$ and $C$} (so also $S-n\la$) {\em ample}, 
so $\chi_i>0$ and $r_i>n\chi_i$.

a) \underline{$\R_i\subset \Sigma_i$:} Here one finds, with $c_i=ns_i+r_iF$, 
again that
$s_i|_{c_i}$ consists of $r_i-n\chi_i$ points and gets the crucial relation
of divisors 
$ns_i|_{c_i}=\sum_{k=1}^{r_i-n\chi_i}F_{u_k}|_{c_i}\sim (r_i-n\chi_i)F|_{c_i}$.

\underline{$\Sigma_i\subset Pfaff_{\C_i}$:} Here one finds the following
(to avoid notational confusion we have denoted the twist parameter in this
set-up by $\mu$; for $\chi_i=1$ everything reduces to (\ref{specialisation}))
\beqa
L|_{c_i}\ox \cO_{c_i}(-F|_{c_i})&=&\left\{ \begin{array}{ll}
\cO_{\E_i}\Big(\frac{1}{2}\Big[ns_i+(r_i+\chi_i-2)F\Big]
+\mu\Big[ns_i-(r_i-n\chi_i)F\Big]\Big)\Big|_{c_i}
& \mbox{for} \; n \, \mbox{even} \\
\cO_{\E_i}\Big((\mu+\frac{1}{2})ns_i
+\Big[(\mu n \chi_i + \frac{\chi_i-2}{2})
-(\mu-\frac{1}{2})r_i\Big]F\Big)\Big|_{c_i}
& \mbox{for} \; n \, \mbox{odd}\end{array} \right.\nonumber\\
\label{specialisation in new set-up}
&\stackrel{\Sigma_i}{\ra}&\left\{ \begin{array}{ll}
\cO_{\E_i}\Big(\frac{n}{2}s_i+\frac{r_i+\chi_i-2}{2}F\Big)\Big|_{c_i}
& \\
\cO_{\E_i}\Big(\Big[r_i-\frac{(n-1)\chi_i+2}{2}\Big]F\Big)\Big|_{c_i}
& 
\end{array} \right.
\eeqa
(here one may compare directly with (\ref{line bundle class}) 
and (\ref{gamma cohomology class});
we did assume here that the case $\la$ even does not occur, so $n$ even
does not allow for a strictly half-integral $\mu$; in the former set-up this
assumption (then: $c_1$ not even)
was fulfilled automatically as there $\chi_i=\C_i\cdot c_1$
was always $1$ from isolatedness). 
One gains again the needed effectivity assertion.

b) In analogy to the reasoning in c) in app.~\ref{proof appendix} 
one gets here (for $n$ even; note $r>n$) that
$h^0(\E, \cO_{\E}(\frac{n}{2}s_i+\frac{r_i+\chi_i-2}{2}F))
=h^0(\C, \cO_{\C}(\frac{r_i+\chi_i-2}{2})
\oplus \bigoplus_{j=2}^{n/2}\cO_{\C}(\frac{r_i+\chi_i-2}{2}-j))
=\frac{r_i+\chi_i-2}{2}+1+\sum_{j=2}^{n/2}(\frac{r_i+\chi_i-2}{2}-j+1)
=\frac{n}{2}(\frac{r_i+\chi_i-2}{2}+1)
-(\frac{1}{2}\frac{n}{2}(\frac{n}{2}+1)-1)$. So one gets for the multiplicity 
\beqa
\label{deviation multiplicity}
k&\geq & \left\{ \begin{array}{ll}
1+ \frac{n}{4}(r_i-\frac{n}{2}+\chi_i-1)
\;\;\;\;\;\;\;\;\;\;\;\;\;\;\;\;\;\;\;\; 
\mbox{for} \;\; n\, \equiv \, 0 \, (2)\\
1+r_i-\frac{(n-1)\chi_i+2}{2}
\;\;\;\;\;\;\;\;\;\;\;\;\;\;\;\;\;\;\;\;\;\;\;\;\;\;\;\;
\mbox{for} \;\; n\, \not\equiv \, 0 \, (2)
\end{array} \right.
\eeqa
So b) of Theorem $Pfaff_{all\, \C\subset X}$
in sect.~\ref{moduli of proj} follows from (\ref{deviation multiplicity}):
choosing $C$ ample in $Y$ makes $S-n\la$ ample in $X$, 
so $r_i> n \chi_i \; \geq \; n$ for $\la$ ample, 
giving again (\ref{r_i bound}); 
the cases $n=3,4,5$
give $k\geq 2$ for all $\C_i$ from (\ref{deviation multiplicity})
(using $r_i-n\chi_i>0$).

\section*{References}
\begin{enumerate}

\item
\label{W}
E. Witten, {\em World-Sheet Corrections Via D-Instantons},
hep-th/9907041, JHEP {\bf 0002} (2000) 030.

\item
\label{FMW}
R. Friedman, J. Morgan and E. Witten, {\em Vector Bundles and F-Theory},
hep-th/9701162, Comm. Math. Phys. {\bf 187} (1997) 679.

\item
\label{BW}
C.~Beasley and E.~Witten,
{\em Residues and World-Sheet Instantons}, hep-th/0304115,
JHEP {\bf 0310} (2003) 065.

\item
\label{SW}
E.~Silverstein and E.~Witten,
{\em "Criteria for Conformal Invariance of (0,2) Models"},
heo-th/9503212, Nucl.Phys. {\bf B444} (1995) 161.

\item
\label{H}
D.~Huybrechts, {\em The Tangent Bundle of a Calabi-Yau Manifold -
Deformations and restriction to Rational Curves},
Comm. Math. Phys. {\bf 171} (1995) 139.

\item
\label{D}
J.~Distler, 
{\em "Resurrecting (2,0) Compactifications"},
Phys.Lett. {\bf B188} (1987) 431;
J.~Distler and B.~Greene, {\em "Aspects of (2,0) String Compactifications"},
Nucl.Phys. {\bf 304} (1988) 1.

\item
\label{BDO}
E.I.~Buchbinder, R.~Donagi and B.A.~Ovrut,
{\em Superpotentials for Vector Bundle Moduli},
hep-th/0205190, Nucl.Phys. {\bf B653} (2003) 400;
{\em Vector Bundle Moduli Superpotentials 
in Heterotic Superstrings and M-Theory},
hep-th/0206203, JHEP {\bf 0207} (2002) 066.

\item
\label{BO}
E.I.~Buchbinder and B.A.~Ovrut,
{\em Vacuum Stability in Heterotic M-Theory},
hep-th/0310112, Phys.Rev. {\bf D69} (2004) 086010.

\item
\label{OPP}
B.A.~Ovrut, T.~Pantev and J.~Park, 
{\em Small Instanton Transitions in Heterotic M-Theory},
hep-th/0001133, JHEP {\bf 0005} (2000) 045.

\item
\label{C1}
G.~Curio, {\em World-sheet Instanton Superpotentials in Heterotic String theory 
and their Moduli Dependence}, arXiv:0810.3087, JHEP {\bf 0909} (2009) 125.

\item
\label{C2}
G.~Curio, {\em Perspectives on Pfaffians of Heterotic World-sheet Instantons},
arXiv:0904.2738, JHEP {\bf 0909} (2009) 131.

\item
\label{C}
G. Curio, {\em Chiral matter and transitions in heterotic string models},
hep-th/9803224, Phys.Lett. {\bf B435} (1998) 39.

\item
\label{CD}
G. Curio and R. Y. Donagi, {\em Moduli in N=1 Heterotic/F-Theory
  Duality}, hep-th/9801057, Nucl.Phys. {\bf B518} (1998) 603.

\end{enumerate}
\end{document}